\documentclass[english,prb,superscriptaddress,preprint]{revtex4-2}
\usepackage[T1]{fontenc}
\usepackage[utf8]{inputenc}
\usepackage{array}
\usepackage{booktabs}
\usepackage{textcomp}
\usepackage{multirow}
\usepackage{amsmath}
\usepackage{stackrel}
\usepackage{graphicx}

\makeatletter

\newcommand{\lyxmathsym}[1]{\ifmmode\begingroup\def\b@ld{bold}
  \text{\ifx\math@version\b@ld\bfseries\fi#1}\endgroup\else#1\fi}

\providecommand{\tabularnewline}{\\}

\makeatother

\usepackage{babel}
\begin{document}
\title{Characterization of Silicon Carbide Biphenylene Network through G$_{0}$W$_{0}$-BSE
Calculations}
\author{Arushi Singh}
\email{arushi.phy@iitb.ac.in}

\affiliation{Department of Physics, Indian Institute of Technology Bombay, Powai,
Mumbai 400076, India}
\author{Vikram Mahamiya}
\email{vmahamiy@ictp.it}

\affiliation{The Abdus Salam International Centre for Theoretical Physics (ICTP),
I-34151 Trieste, Italy}
\author{Alok Shukla}
\email{shukla@iitb.ac.in}

\affiliation{Department of Physics, Indian Institute of Technology Bombay, Powai,
Mumbai 400076, India}
\begin{abstract}
Two-dimensional silicon carbide stands out among 2D materials, primarily
due to its notable band gap, unlike its carbon-based counterparts.
However, the binary nature and non-layered structure of bulk SiC present
challenges in fabricating its 2D counterpart. Recent advancements
in technology have led to the successful synthesis of atomically thin,
large-scale epitaxial monolayers of hexagonal-SiC {[}Polley \emph{et
al.},\emph{ }Phys. Rev. Lett. 130, 076203 (2023){]} and Si$_{9}$C$_{15}$
{[}Gao \emph{et al.}, Adv. Mater. 34, 2204779 (2022){]}, marking a
significant milestone in semiconductor research. Inspired by these
advancements, we have computationally designed another stable phase
of 2D-SiC in the popular biphenylene network, termed SiC-biphenylene.
This structure is characterized by interconnected polygons of octagons,
hexagons, and tetragons arranged periodically. The dynamical and thermal
stability has been confirmed through \emph{ab initio} phonon dispersion
and molecular dynamics simulations. The structure demonstrates a high
melting point of approximately 3475 K and a ``direct'' band gap
of 2.16 eV using the HSE06 functional. Upon considering many-body
effects, the quasiparticle band gap widens to 2.89 eV at the G$_{0}$W$_{0}$
level, indicating pronounced electron correlation effects within the
material. Our analysis further reveals that the effective mass of
charge carriers exhibits higher values along the $\Gamma\rightarrow Y$
compared to the $\Gamma\rightarrow X$ direction. Moreover, the optical
spectrum obtained from solving the Bethe-Salpeter equation (G$_{0}$W$_{0}$+BSE)
identifies the first optically active exciton peak at 2.07 eV, corresponding
to a strongly bound exciton with a binding energy of 0.82 eV. The
effective mass and Bohr radius of the exciton are calculated to be
1.01 $m_{0}$ and 2.14 $\text{Å}$, respectively, demonstrating the
characteristic of a Frenkel exciton. Furthermore, the investigation
into stable bilayer structures across various stacking configurations
(AA, AA', and AB-stacked) highlights the impact of stacking patterns
on excitonic binding energies, with AA-stacked configuration indicating
the presence of Mott-Wannier exciton. Our investigation extends to
identifying the stable bulk phase of SiC-biphenylene, revealing lower
self-energy corrections compared to monolayer and bilayer structures,
attributed to increased electron delocalization in bulk structures.
\end{abstract}
\maketitle

\section{Introduction\label{sec:introd}}

The diverse hybridization $(sp,sp^{2},sp^{3})$ schemes of the valence
electronic orbitals of the carbon atom lead to an extensive array
of one- (1D), two- (2D), and three- (3D) dimensional allotropic forms
of carbon, with controllable physicochemical properties and potential
application in various fields. A few carbon phases, including fullerene
\citep{Kroto1985}, graphene \citep{science.1102896}, graphdiyne
\citep{li2010architecture}, phagraphene \citep{wang2015phagraphene},
naphyne \citep{li2020architecture}, graphtetrayne \citep{pan2021direct},
have been successfully synthesized in laboratory experiments. Numerous
carbon allotropes such as, pentaheptite \citep{crespi1996prediction},
haeckelites \citep{PhysRevLett.84.1716}, graphenylene \citep{song2013graphenylene},
pentahexoctite \citep{sharma2014pentahexoctite}, $\varPsi$-graphene
\citep{li2017psi}, graphyne \citep{li2019new}, tetra-graphene \citep{bandyopadhyay2020review},
D-graphyne \citep{zhang2021semimetallic}, P-graphyne \citep{zhang2021semimetallic},
etc., have been identified theoretically. Hence, an extensive experimental
and theoretical investigations have been conducted on carbon nanomaterials.
To date, the SACADA database \citep{SACADA} currently catalogs around
500 different carbon allotropes, highlighting the extensive diversity
present within the carbon family and its significance in the field
of materials science.

In 2010, Hudspeth \emph{et al}. \citep{Hudspeth} theoretically predicted
a novel fully $sp^{2}$ bonded planar carbon allotrope known as the
biphenylene network, which consists of the periodically connected
octagon\emph{- (o)},\emph{ }hexagon\emph{- (h)},\emph{ }and square-
\emph{(s)} rings. The carbon atoms in the biphenylene network are
three-fold coordinated, thereby maintaining the same number of neighbors
as in the hexagonal network of graphene structure. However, the associated
$sp^{2}$ bonds of carbon atoms undergo deformation, transitioning
to a purely metallic state in biphenylene \citep{bafekry2021biphenylene}.
The non-benzoid octagonal rings of biphenylene facilitate enhanced
lithium-ion and sodium-ion storage capacity compared to graphene,
thus positioning it as a promising candidate for application as an
anode material \citep{D2CP00798C,ferguson2017biphenylene,D2CP04752G}.
A first-principles investigation by Liu \emph{et al.} \citep{Liu2021}
revealed that the carbon atoms in the tetragonal rings of biphenylene
are substantially positively charged, rendering them viable active
sites for the oxygen reduction reaction (ORR). The rotational vibration
phonon mode of the six-membered carbon ring exhibits strong coupling
with electrons, leading to the possibility of superconductivity with
a predicted critical temperature of 6.2 K \citep{ge2021superconductivity}.
This critical temperature can further be enhanced to 27.4 K and 21.5
K by applying small uniaxial strain and impurity doping, respectively.
The biphenylene structure can also generate quasi-1D nanotubes and
nanoribbons exhibiting diverse edge geometries compared to graphene
\citep{PhysRevB.105.035408}. In a recent breakthrough, Fan \emph{et
al. }\citep{fan2021biphenylene} successfully fabricated the ultra-flat
biphenylene nanoribbons on Au(111) surface through an interpolymer
dehydrofluorination (HF-zipping) reaction. Scanning probe microscopy
(SPM) analysis was conducted to investigate the influence of nanoribbon
widths on their band gaps. It was established that metallicity emerges
with increasing nanoribbon width, thus affirming the metallic properties
of the biphenylene sheet, consistent with previous theoretical prediction
\citep{bafekry2021biphenylene,luder2015electronic,Hudspeth}. This
marks a significant milestone in the experimental realization of a
nanostructure composed of non-benzoid rings and their distinct characteristics.

Two-dimensional silicon carbide (2D-SiC) monolayers represent a novel
category of semiconducting monolayers, distinguished by their exceptional
properties. While the theoretical concept of a two-dimensional allotrope
of SiC has been under discussion for an extended period, its experimental
synthesis has remained elusive owing to the non-layered structure
of bulk SiC. Unlike graphene, the bulk structure of SiC cannot serve
as a parent structure for the mechanical exfoliation to create 2D
materials. However, recent progress in synthesis methodologies has
introduced a novel approach for generating these non-layered 2D-structures
through epitaxial growth on appropriate substrates. Very recently,
Polley et al. \citep{PhysRevLett.130.076203} reported the large-scale
synthesis of an epitaxial monolayer of honeycomb SiC through the annealing
of thin films of transition metal carbides grown on 4H-SiC(0001) substrate.
This newly synthesized 2D phase of SiC exhibits a planar structure
and possesses a wide band gap of 2.5 eV. It demonstrates excellent
stability at high temperatures, up to 1200 °C in a vacuum environment.
Another significant milestone in 2D-SiC research is the recent achievement
by Gao and co-workers \citep{adma.202204779}, who synthesized large-scale
atomic monolayer Si$_{9}$C$_{15}$ on Ru(0001) and Rh(111) substrates.
This synthesis involves growing a graphene layer on a Ru or Rh substrate,
followed by silicon evaporation onto the graphene surface and high-temperature
annealing to form the Si$_{9}$C$_{15}$ layer. The resulting monolayer
Si$_{9}$C$_{15}$ exhibits robust environmental stability and features
a buckled honeycomb structure with a bandgap of \ensuremath{\approx}
1.9 eV. 

Inspired by the recent experimental synthesis of 2D-SiC allotropes,
namely hexagonal SiC \citep{PhysRevLett.130.076203} and Si$_{9}$C$_{15}$
\citep{adma.202204779}, alongside the established synthesis of the
biphenylene network utilizing carbon atoms \citep{fan2021biphenylene},
in this work we have computationally designed the silicon carbide
counterpart within the biphenylene network, referred to as SiC-biphenylene.
Several novel materials have been foreseen through computational modeling,
which emphasize the crucial contribution of theoretical predictions
in advancing the development of novel materials. In recent investigations
involving both computational predictions and laboratory synthesis,
a variety of 2D silicon carbide (SiC) materials have surfaced that
share structural topology with carbon-based monolayers \citep{Long_2021,VANHOANG2019236,C9NR08755A,WEI2021159201}.
SiC-based monolayers exhibit a wide band gap with high thermal and
mechanical stability, making them useful for high-power electronic
devices with distinct on/off switching behavior \citep{Susi2017,PhysRevB.80.155453,PhysRevB.108.235311,Lin2012}.
The recent experimental fabrication of a 2D biphenylene network of
carbon atoms and 2D-SiC monolayer strongly suggest the potential realization
of the SiC-biphenylene structure in the near future. 

The remainder of this paper is structured as follows. The subsequent
Sec. \ref{sec:Method} outlines the computational details and theoretical
methods, including descriptions of the ground-state, quasiparticle,
and electron-hole interaction calculations. In Sec. \ref{subsec:A},
we present a comprehensive analysis of the dynamical, thermal, and
mechanical stability of the novel SiC-biphenylene monolayer structure.
Sec. \ref{subsec:B} explores the electronic and excitonic properties
in detail, while Sec. \ref{subsec:c} investigates the mechanical
strength and elasticity of the structure. In Sec. \ref{subsec:d},
we discuss the results of finite-temperature molecular dynamics simulations
of the monolayer over a wide temperature range. Additionally, in Sec.
\ref{subsec:e}, we explore the stability of SiC-biphenylene in various
bilayer stacking configurations and its bulk structure, providing
a comprehensive summary of the material’s transition from monolayer
to bilayer to and finally to the bulk forms in Sec. \ref{sec:conclusion}.

\section{method \label{sec:Method}}

We performed the first principles calculations using density functional
theory (DFT) employing the VASP simulation package \citep{PhysRevB.54.11169,PhysRevB.59.1758,KRESSE199615}.
Electron-ion interactions were described using the projector-augmented
plane wave (PAW) approximation \citep{PhysRevB.50.17953}. Electronic
wave functions were expanded using a plane wave basis set with a kinetic
energy cutoff of 520 eV. Monkhorst-Pack k-point grid of $9\times9\times1$
has been employed to sample the Brillouin zone \citep{PhysRevB.13.5188}.
Additionally, a vacuum space of 20 Å was maintained along the z-direction
to prevent interactions between periodic images of the monolayer.
The energy and force convergence limit is set to be $10^{-7}$eV and
$10^{-3}$eV$\text{Å}^{-1}$, respectively. The structural relaxations
and the ground state energies were calculated using the generalized
gradient approximation (GGA) within the Perdew\textminus Burke\textminus Ernzerhof
(PBE) parametrization \citep{PhysRevLett.77.3865}. It is widely recognized
that the GGA functional tends to underestimate the band gap, while
the Hatree-Fock (HF) method overestimates it \citep{RevModPhys.80.3}.
Therefore, to get a more accurate prediction of band gap \citep{Barone2011},
we employed screened hybrid-functional HSE06 \citep{10.1063/1.1564060,10.1063/1.2204597,10.1063/1.1760074},
which is constructed by mixing 25\% of Fock exchange $(E_{x}^{HF})$
with 75 \% of PBE exchange $(E_{x}^{PBE})$ in the short range. Thus,
the exchange-correlation energy is calculated as:

\begin{equation}
E_{xc}^{HSE06}=\frac{1}{4}E_{x}^{HF,SR}(\omega)+\frac{3}{4}E_{x}^{PBE,SR}(\omega)+E_{x}^{PBE,LR}(\omega)+E_{c}^{PBE},\label{eq:eqn-1}
\end{equation}

where $E_{x}^{HF,SR}$ and $E_{x}^{PBE,SR}$ denote the short-range
components of Hartree-Fock and PBE exchange energies, respectively.
$E_{x}^{PBE,LR}$ is the long-range PBE exchange energy, and $E_{c}^{PBE}$
is the PBE correlation energy. $\omega=0.11$ bohr$^{-1}$ serves
as the screening parameter \citep{10.1063/1.2404663}.

The lattice dynamical stability is verified by evaluating the phonon
dispersion using the Density Functional Perturbation Theory (DFPT)
approach. For this evaluation, we employed a $9\times9\times1$ Monkhorst-Pack
k-point grid for the monolayer structure and a $5\times5\times1$
Monkhorst-Pack k-point grid for both the bilayer and bulk structures.
These grids were selected to ensure adequate sampling of the Brillouin
zone while balancing computational cost and accuracy. The PHONOPY
\citep{Togo_2023} software is employed to determine the requisite
force constants essential for calculating the phonon spectra. The
thermal stability at finite temperature is tested by employing ab-initio
molecular dynamics (AIMD) simulations \citep{nose1984molecular}.
First, the system is kept in a microcanonical ensemble (NVE) for 6
ps with 1 fs time steps while raising the temperature to a desired
value. Then, the system is equilibrated in a canonical ensemble (NVT)
for 6 ps at that temperature, utilizing a Nose--Hoover thermostat
\citep{10.1063/1.447334}. 

In analyzing the energetics, the cohesive energy $(E_{C})$ and formation
energy $(\triangle E_{F})$ of the monolayer system have been calculated
using the following expressions :

\begin{equation}
E_{C}=\frac{E_{T}^{biphenylene-SiC}-(N_{Si}E_{T}^{Si}+N_{C}E_{T}^{C})}{N_{Si}+N_{C}},\label{eq:eqn-2}
\end{equation}

\begin{equation}
\triangle E_{F}=\frac{\mu^{biphenylene-SiC}-(N_{Si}\mu^{Si}+N_{C}\mu^{C})}{N_{Si}+N_{C}}\label{eq:eqn-3}
\end{equation}

where $E_{T}^{biphenylene-SiC}$represents the total energy of the
SiC-biphenylene unit cell, $E_{T}^{SiC}$and $E_{T}^{C}$ denote the
total energies of isolated Si and C atoms, respectively. $\mu^{biphenylene-SiC}$
stands for the chemical potential of the SiC-biphenylene unit cell;
$\mu^{Si}$ and $\mu^{C}$ represent the chemical potentials of the
stable bulk phases of Si and C atoms, respectively. $N_{Si}$ and
$N_{C}$ refer to the number of Si and C atoms in a monolayer unit
cell, respectively.

To model the interlayer interactions in bilayer systems, we employed
the semi-empirical DFT-D3 method developed by Grimme \citep{10.1063/1.3382344,https://doi.org/10.1002/jcc.21759}.
In this approach, van der Waals interactions are incorporated by adding
a semi-empirical dispersion potential to the conventional DFT energy
framework. This dispersion potential is parameterized using empirical
fit parameters derived from a range of reference systems. To more
accurately capture long-range van der Waals (vdW) forces, we also
employed non-local vdW functionals, namely optB86b-vdW \citep{PhysRevB.83.195131}
and optB88-vdW \citep{Klime_2010}. These functionals account for
vdW interactions directly within the exchange-correlation functional,
rather than relying on an empirical correction. The inter-layer binding
energy $E_{ib}$ is then calculated using the expression :

\begin{equation}
E_{ib}=\frac{E_{T}^{bilayer-biphenylene-SiC}-2E_{T}^{biphenylene-SiC}}{N_{Si}+N_{C}},\label{eq:eqn-4}
\end{equation}

where $E_{T}^{bilayer-biphenylene-SiC}$ and $E_{T}^{biphenylene-SiC}$
represent the total energies of the bilayer and monolayer systems,
respectively, calculated using the DFT-D3 corrections. $N_{Si}$ and
$N_{C}$ denote the number of Si and C atoms, respectively, in a bilayer
unit cell.

While the DFT approach has demonstrated remarkable efficacy in characterizing
the ground-state properties of materials, its ability to adequately
describe excited-state phenomena, such as optical properties, in 2D
systems is quite limited due to the strong excitonic effects inherent
to these materials. Therefore, we employed many-body perturbation
theory (MBPT) approaches to accurately calculate and understand the
many-electron properties of these systems \citep{Strinati1988,PhysRevLett.62.1169}.
We computed the quasi-particle (QP) band structure using single shot
G$_{0}$W$_{0}$ approximation \citep{RevModPhys.74.601}, utilizing
the eigenvalues and eigenstates from HSE06 as input to determine the
QP energies. This method approximates the quasi-particle (QP) self-energy
as the product of the Green's function G and the dynamically screened
Coulomb interaction W, serving as a first-order perturbation correction
to the Kohn-Sham eigenvalues. In practice, QP energies are derived
as 

\begin{equation}
E_{n\boldsymbol{k}}^{qp}=\epsilon_{n\boldsymbol{k}}+Z_{n\boldsymbol{k}}[\varSigma_{n\boldsymbol{k}}(G,W;\epsilon_{n\boldsymbol{k}})-\upsilon_{n\boldsymbol{k}}^{xc}]\label{eq:eqn-5(a)}
\end{equation}

where $\epsilon_{n\boldsymbol{k}}$ represents the Kohn-Sham eigenvalues.
$\varSigma_{n\boldsymbol{k}}(\omega)$ and $\upsilon_{n\boldsymbol{k}}^{xc}$
denote the expectation values of the self-energy and the exchange-correlation
potential, respectively, over the n\textbf{k} Kohn-Sham eigenvector.
$Z_{n\boldsymbol{k}}$ signifies the quasiparticle (QP) renormalization
factor, which is defined as

\begin{equation}
Z_{n\boldsymbol{k}}=\left[1-\left.\frac{\partial\varSigma_{n\boldsymbol{k}}(\omega)}{\partial\omega}\right|_{\omega=\epsilon_{n\boldsymbol{k}}}\right]^{-1}\label{eq:eqn-5(b)}
\end{equation}

This approach allows the calculation of the quasiparticle shifts for
all electronic states and at all considered \textbf{k} points. We
employed 160 empty conduction band and $7\times7\times1$ k-mesh grid
to conduct the G$_{0}$W$_{0}$ calculations. Subsequently, the QP
band structure was interpolated using the maximally localized Wannier
functions (MLWFs) \citep{RevModPhys.84.1419} implemented in the WANNIER90
package \citep{MOSTOFI20142309}.

The excitonic optical properties, which incorporate electron-hole
interactions, were investigated by solving the Bethe-Salpeter equation
(BSE) \citep{PhysRevB.62.4927,PhysRevB.78.205108} built upon G$_{0}$W$_{0}$
eigenvalues and wave functions. The excitation energy $\Omega^{S}$
and the corresponding exciton ampltiude $A_{vc\boldsymbol{k}}^{S}$
of the correlated electron-hole excitations are determined by solving
the following BSE equation

\begin{equation}
(E_{c\boldsymbol{k}}-E_{v\boldsymbol{k}})A_{vc\boldsymbol{k}}^{S}+\sum_{v'c'\boldsymbol{k}'}\left\langle vc\boldsymbol{k}|K^{eh}|v'c'\boldsymbol{k}'\right\rangle A_{v'c'\boldsymbol{k}'}^{S}=\Omega^{S}A_{vc\boldsymbol{k}}^{S}\label{eq:eqn-5(c)}
\end{equation}

Here, $E_{c\boldsymbol{k}}$ and $E_{v\boldsymbol{k}}$ are the G$_{0}$W$_{0}$
eigenvalues of conduction and valence bands, respectively, at a specific
\textbf{k }point. $K^{eh}$ is the kernel describing the screened
interaction between excited electrons and holes. The excitonic eigenstates
were constructed based on the 48 highest valence bands and the 24
lowest conduction bands.

The Bohr radius $a^{exciton}$ of a ground state exciton with a corresponding
binding energy $E_{b}^{exciton}$ can be calculated using the following
equation \citep{GODET2001168,SHAHROKHI20171185,SHAHROKHI2016377}

\begin{equation}
a^{exciton}=\frac{\varepsilon_{r}}{\mu^{exciton}}\times a_{h}\times m_{0}\label{eq:eqn-5(d)}
\end{equation}

Here, $m_{0}$ represents the rest mass of the electron, $\varepsilon_{r}$
denotes the dielectric constant, $a_{h}$ stands for the Bohr radius
$(0.529\text{Å})$, $R_{h}$ signifies the Rydberg constant (13.6
eV). The effective mass of the exciton, denoted by $\mu^{exciton}$,
is determined as $\mu^{exciton}=\frac{E_{b}^{exciton}}{R_{h}}\times\varepsilon_{r}^{2}\times m_{0}$

. 

The mechanical properties were calculated by subjecting the equilibrium
structure to strains ranging from -2\% to +2\% in increments of 0.05\%.
The elastic energy $\triangle E(V,\{\varepsilon_{i}\})$ of a material
under strain is expressed using Voigt notation \citep{PhysRevB.85.125428}
and can be described in the following harmonic approximation:
\begin{equation}
\triangle E(V,\{\varepsilon_{i}\})=E(V,\{\varepsilon_{i}\})-E(V_{0},0)=\frac{V_{0}}{2}\stackrel[i,j=1]{6}{\sum}C_{ij}\varepsilon_{i}\varepsilon_{j}\label{eq:eqn-6}
\end{equation}

where $E(V,\{\varepsilon_{i}\})$ and $E(V_{0},0)$ are the total
energies of the strained and unstrained lattice, with the volumes
of $V$ and $V_{0}$, respectively. Thus, the elastic tensor $(C_{ij})$
is derived from the second-order derivative of the total energies
versus strain. The orientation-dependent Young's Modulus $E(\theta)$
and the Poisson's ratio $\nu(\theta)$ can then further be obtained
using the following equations :

\begin{equation}
E(\theta)=\frac{C_{11}C_{22}-C_{12}^{2}}{C_{11}\sin^{4}(\theta)+A\sin^{2}(\theta)\cos^{2}(\theta)+C_{22}\cos^{4}(\theta)},\label{eq:eqn-7}
\end{equation}

\begin{equation}
\nu(\theta)=\frac{C_{12}\sin^{4}(\theta)-B\sin^{2}(\theta)\cos^{2}(\theta)+C_{12}\cos^{4}(\theta)}{C_{11}\sin^{4}(\theta)+A\sin^{2}(\theta)\cos^{2}(\theta)+C_{22}\cos^{4}(\theta)}\label{eq:eqn-8}
\end{equation}

where $A=\frac{C_{11}C_{22}-C_{12}^{2}}{C_{66}}-2C_{12}$ and $B=\frac{C_{11}C_{22}-C_{12}^{2}}{C_{66}}-2C_{12}$. 

\section{Results and analysis \label{sec:Results-and-analysis}}

\subsection{Structural conformation and experimental feasibility\label{subsec:A}}

\begin{figure}[h]
\includegraphics[clip,scale=0.65]{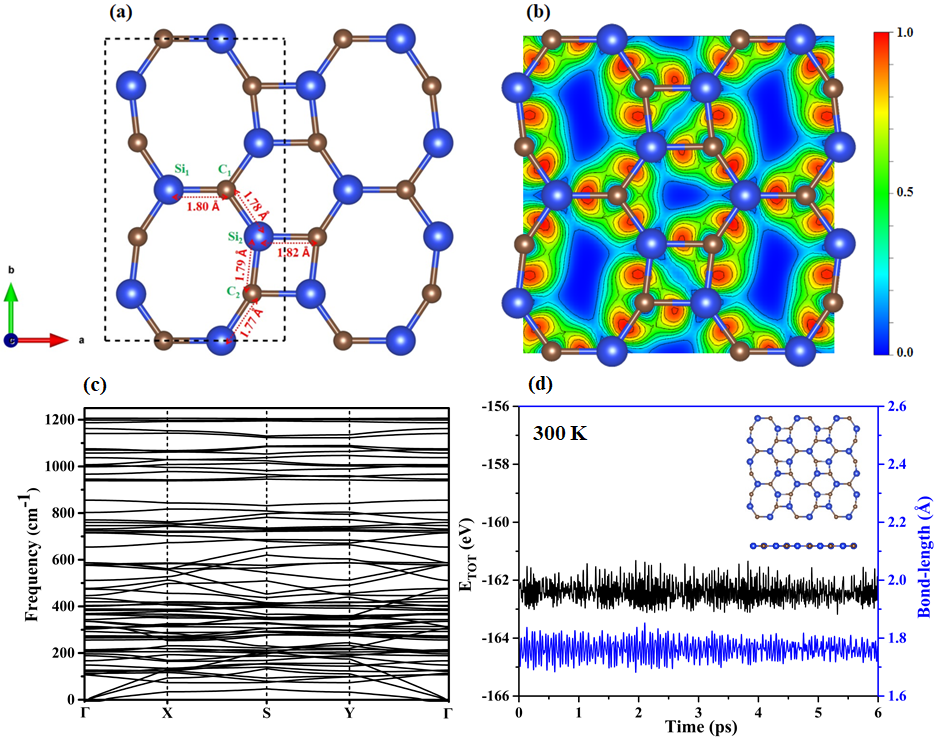}\caption{\label{fig:figure-01} (a) Optimized atomic structure of 2D SiC-biphenylene
monolayer. The unit cell is delineated by black dashed lines. Two
distinct atomic sites are represented by (Si$_{1}$, C$_{1}$) and
(Si$_{2}$, C$_{2}$); (b) Contour line diagrams of the ELF electronic
distributions; (c) Calculated phonon modes with all real frequencies
along the high symmetry directions of Brillouin zone; (d) Fluctuations
in total energy and bond length at 300 K for 6 ps with the steps of
1 fs (inset shows the optimised structure at 300 K)}
\end{figure}

This section presents the atomic arrangement and structural properties
of an optimized ground state geometry of SiC-biphenylene monolayer
(Fig.\ref{fig:figure-01}). The lattice belongs to the Pmma space
group and possesses an orthorhombic crystal structure. The regular
primitive cell, composed of 6-Si-atoms and 6-C-atoms, consists of
two fused octagons (enclosed by black dashed lines in Fig.\ref{fig:figure-01}a)
with lattice vectors $\vec{a}$ and $\vec{b}$ of magnitudes 5.61
$\text{Å}$ and 9.44 $\text{Å}$, respectively. The $b/a$ ratio
is 1.68, which corresponds to structural anisotropy along x- and y-direction,
giving rise to direction-dependent electronic, mechanical and, optical
properties, which we shall discuss later in Sec.\ref{subsec:B} \&
\ref{subsec:c}. The planar structure of SiC-biphenylene monolayer
consists of eight-, six-, and four-membered polygons of connected
octagon, hexagon and tetragon. As illustrated in Fig.(\ref{fig:figure-01}a),
the structure consists of two types of atomic sites corresponding
to Si (namely, Si$_{1}$ and Si$_{2}$) and C (namely, C$_{1}$ and
C$_{2}$) atoms, respectively, where atoms Si$_{1}$ and C$_{1}$
are shared by the two adjacent octagons and one hexagon, while atoms
Si$_{2}$ and C$_{2}$ are shared by an adjacent octagon, hexagon
and tetragon. This configuration results in four different bond lengths:
$\overline{Si_{1}C_{1}}=1.80\text{ Å}$, $\overline{Si_{2}C_{1}}=1.78\text{ Å}$,
$\overline{Si_{2}C_{2}}=1.79\text{ Å}$ $(1.82\text{ Å})$, $\overline{Si_{1}C_{2}}=1.77\text{ Å}$,
as shown in Fig.(\ref{fig:figure-01}a). The Bader charge analysis
also confirmed that the two types of atomic sites ( i.e. (Si$_{1}$,
C$_{1}$) and (Si$_{2}$,C$_{2}$) ) correspond to different effective
charges, and hence exhibit different contribution to the energy bands
(Fig.\ref{fig:figure-02}b \& c) and the site-dependent chemical activity
against the foreign atom. To validate our DFT schemes employed, we
optimized the geometry of a recently synthesized novel biphenylene
carbon sheet, and the lattice vectors are calculated to be 4.52 $\text{Å}$
and 3.76 $\text{Å}$, which match precisely with previous theoretical
reports \citep{bafekry2021biphenylene,D1NR07959J,D1TC04154A,VEERAVENKATA2021893,Luo2024}.
Further, to reveal the bonding character between the constituent atoms,
the Electron Localization Function (ELF) is computed (Fig.\ref{fig:figure-01}b),
and the values are renormalized between 0 (completely delocalized
electrons) to 1 (completely localized electrons). A highly pronounced
localized electron density is found between Si and C atoms, indicating
the strong covalent-type character of the bond between them. A small
amount of delocalized electrons around the C-atom is observed, which
is attributed to its higher electronegativity than the Si-atom, leading
to the slight deformation of the octagon and tetragon rings as compared
to the biphenylene structure composed of all carbon atoms. 

Next, we addressed the experimental feasibility of the freestanding
monolayer of SiC-biphenylene based on cohesive energy, formation energy,
phonon spectrum, ab-initio molecular dynamics (AIMD) simulations,
and elastic constants. The computed cohesive energy $E_{C}$ of SiC-biphenylene
is -5.72 eV/atom which is lower than that of biphenylene (-7.40 eV/atom)\citep{luo2021first}
but comparable to those of biphenylene-MoS$_{2}$(-4.94 eV/atom) \citep{gorkan2023can}
and SiC monolayer (-5.93 eV/atom) \citep{PhysRevB.108.235311,PhysRevB.81.075433},
demonstrating the structural stability of the the system. The formation
energy $\Delta E_{F}$ is found to be 0.52 eV/atom, which is close
to the formation energy of experimentally synthesized materials, such
as SiC monolayer (0.29 eV/atom) \citep{PhysRevB.102.134103,PhysRevB.108.235311}
and biphenylene (0.46 eV) (see Table \ref{tab:tab-1}), which suggests
that the monolayer structure of SiC-biphenylene is energetically stable.
Further, we computed the elastic constants which satisfy the Born--Huang
criteria \citep{born1955dynamical} $C_{11}C_{22}-C_{12}^{2}>0$ and
$C_{66}>0$ showcasing the mechanical stability of the newly designed
monolayer system (detailed results are discussed in Sec.\ref{subsec:c}).
To evaluate the lattice dynamical stability at 0 K, phonon dispersion
calculations were conducted (Fig.\ref{fig:figure-01}c). The highest
optical phonon mode reaches a frequency of 1206.41 $cm^{-1}$. The
three acoustic phonon modes exhibit slight imaginary frequencies near
the $\Gamma$-point, which, in all likelihood, are numerical artifacts.
These artifacts typically arise from limitations in the computational
method, such as the use of finite-size supercells or insufficient
convergence settings, particularly affecting long-wavelength (low-energy)
acoustic phonons, and have been frequently reported in the literature
\citep{PhysRevB.71.205214,PhysRevX.7.021019}. The absence of large
imaginary frequencies in the overall phonon dispersion confirms the
dynamic stability of the structure at 0 K. Additionally, the material's
response to the thermal perturbations at room temperature (300 K)
was also assessed by employing \emph{ab-initio} Molecular Dynamics
(AIMD) Simulations. The total energy exhibited a maximum fluctuation
of 1.08 eV, indicating a stable thermal response (Fig.\ref{fig:figure-01}d).
Furthermore, bond length fluctuations were analyzed, with the $\overline{Si_{2}C_{1}}$
bond showing a variation of 0.09 Å (Fig.\ref{fig:figure-01}d). Across
the monolayer, which contains four different types of bond lengths,
the fluctuations ranged from 0.07 Å to 0.12 Å. These small variations
in bond lengths suggest that the system retains structural integrity
under thermal perturbation, confirming its thermodynamic stability
at the room temperature. The snapshot of the final stage of AIMD simulation
(inset of Fig.\ref{fig:figure-01}d), demonstrates that the structure
is stable without any bond-reconstruction. Thus, our simulation results
collectively suggest the excellent structural, thermal, and mechanical
stability of the newly predicted SiC-biphenylene monolayer system. 

\begin{table}[h]
\begin{tabular}{>{\centering}p{3cm}>{\centering}p{3cm}>{\centering}p{3.5cm}>{\centering}p{3.5cm}>{\centering}p{3.5cm}}
\toprule 
Material & Lattice Constants

(Å) & Formation Energy

$\Delta E_{F}$ (eV/atom) & Cohesive Energy

$E_{C}$ (eV/atom) & Band Gap

$E_{g}^{HSE06}$ / $E_{g}^{G_{0}W_{0}}$

(eV)\tabularnewline
\midrule
\midrule 
SiC-biphenylene  & 5.61, 9.44  & 0.52 & -5.72 & 2.16 / 2.89\tabularnewline
\midrule 
Biphenylene \citep{luo2021first,Luo2024} & 4.52, 3.76  & 0.46 & -7.40  & Metallic\tabularnewline
\midrule 
h-SiC \citep{PhysRevB.108.235311,PhysRevB.81.075433,C5CP00601E} & 3.09, 3.09 & 0.29 & -5.94 & 3.38 / 3.96\tabularnewline
\bottomrule
\end{tabular}\caption{\label{tab:tab-1} Calculated values of lattice constants, formation
energy ($\Delta E_{F}$), cohesive energy ($E_{C}$ ) and band gap
($E_{g}^{HSE06}$ / $E_{g}^{G_{0}W_{0}}$) for 2D SiC-biphenylene
monolayer. Corresponding values of 2D Biphenylene and h-SiC monolayers
have also been included for the sake of comaprison. }

\end{table}

\subsection{Electronic and excitonic properties \label{subsec:B}}

\begin{figure}[h]
\includegraphics[clip,scale=0.5]{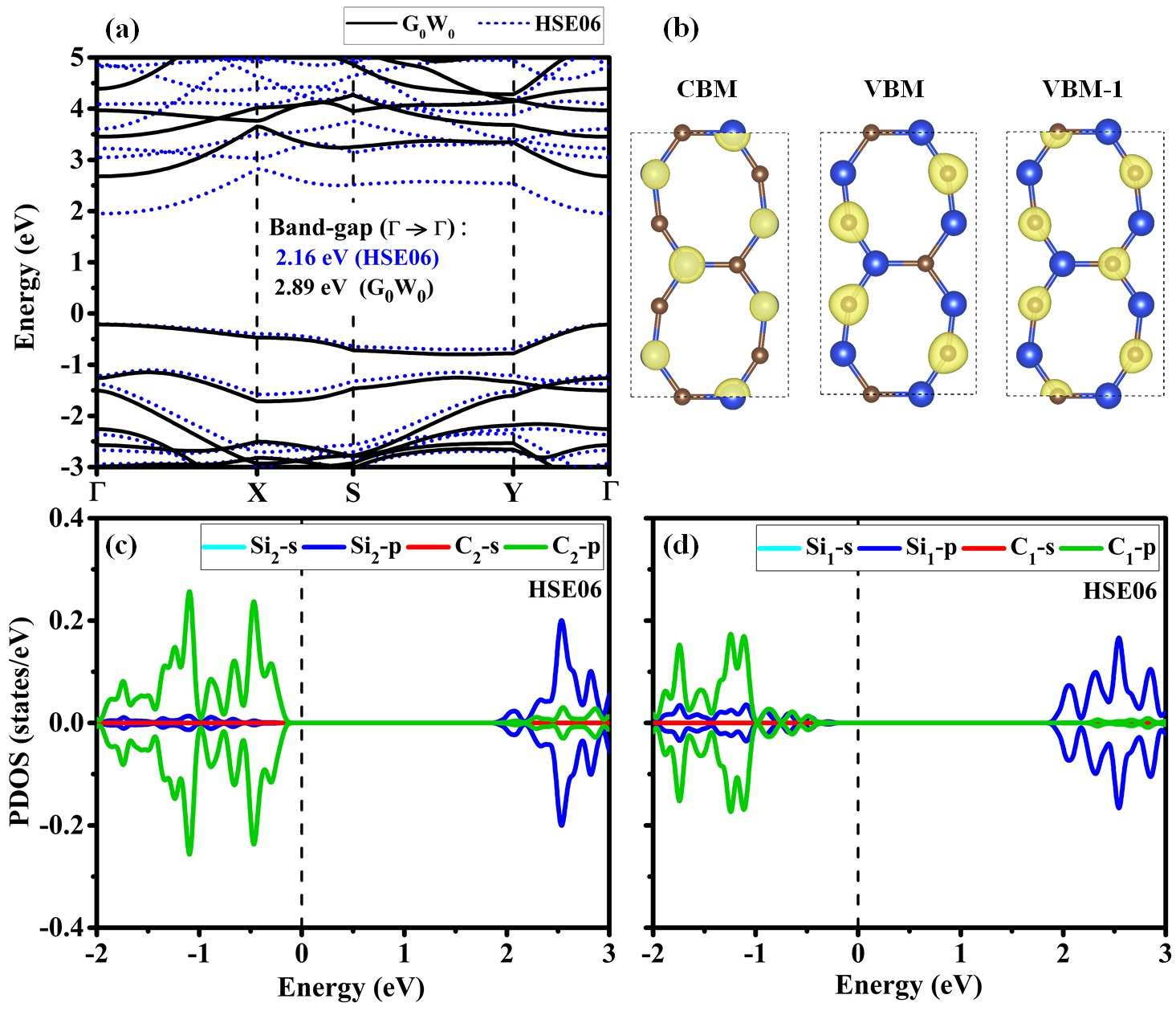}

\caption{\label{fig:figure-02} (a) Electronic energy band structure at HSE06
(blue dotted line) and G$_{0}$W$_{0}$ (black solid line) levels
of theory; (b) Partial charge density plot for the lower conduction
band (CBM) and the upper two valence bands (VBM and VBM-1); (c) \&
(d) Projected density of states (PDOS) of a 2D SiC-biphenylene monolayer.
Zero of the energy is set to E$_{F}$.}
\end{figure}

The electronic band structure of a monolayer of SiC-biphenylene has
been investigated using HSE06 and G$_{0}$W$_{0}$ methodologies,
as depicted in Fig. \ref{fig:figure-02}a. In this figure, the blue
dotted line represents the band structure obtained from HSE06, while
the black line corresponds to the band structure derived from G$_{0}$W$_{0}$
calculations. The SiC-biphenylene monolayer exhibits a ``direct''
band gap of 2.16 eV at the HSE06 level, with the Valence Band Maximum
(VBM) and Conduction Band Minimum (CBM) located at the high-symmetry
point $\varGamma$ of the Brillouin zone. Upon considering many-body
effects, the Quasiparticle (QP) band gap increases to 2.89 eV at the
G$_{0}$W$_{0}$ level, indicating a self-energy correction of 0.73
eV. This substantial quasiparticle correction arises from the enhanced
$e-e$ interactions and reduced screening in lower dimensions. Though
HSE06 incorporates non-local exchange effects, it results in a smaller
gap compared to the G$_{0}$W$_{0}$ method. This underscores the
significant influence of the Coulomb screening effect, which is predominantly
captured by the self-energy operator in the G$_{0}$W$_{0}$ scheme,
on the electronic structure of monolayer SiC-biphenylene. However,
the general shape and nature of band gap remain consistent across
both HSE06 and G$_{0}$W$_{0}$ methodologies.

Fig. \ref{fig:figure-02}b displays the partial charge density of
the Conduction Band Minimum (CBM) and two uppermost Valence Bands
(i.e., VBM \& VBM-1). The primary contribution to the VBM charge density
comes from electrons distributed around the C$_{2}$ type atoms, while
electrons from both C$_{1}$ and C$_{2}$ atoms contribute to the
VBM-1. This inference is further supported by PDOS (Fig. \ref{fig:figure-02}c-d),
which indicates that the C$_{2}$ atoms dominate the valence band
edge compared to the C$_{1}$ atoms. Similarly, in the conduction
band, the contribution of the Si$_{1}$ atoms surpasses that of the
Si$_{2}$ atoms. This highlights the site-specific chemical reactivity
characteristic of the Si and C atoms. The valence band edge primarily
comprises the p-orbitals of the carbon atoms, while the p-orbital
of the Si atom primarily contributes to the conduction band edge.
As the band edges comprise different atoms, this property facilitates
the separation of charge carriers and extends the lifetime of excitons,
suggesting potential applications as a photoactive layer in solar
cells. Additionally, the symmetric nature of the PDOS for spin-up
and spin-down channels confirms the non-magnetic nature of this material.
Therefore, we employed non spin-polarized calculations for further
exploration.

The mobility and behavior of charge carriers in semiconductors are
heavily influenced by the effective masses of electrons and holes,
which determine their usefulness in optoelectronic devices. Notably,
there exists an anisotropy in the effective masses of electrons and
holes around the Conduction Band Minimum (CBM) and Valence Band Maximum
(VBM) respectively. The electron effective mass is estimated to be
0.25 m$_{0}$ and 1.34 m$_{0}$ along $\Gamma\rightarrow Y$ and $\Gamma\rightarrow X$
directions, respectively. While the effective mass of holes is determined
to be 0.28 m$_{0}$ and 2.21 m$_{0}$ along $\Gamma\rightarrow Y$
and $\Gamma\rightarrow X$ directions, respectively. Consequently,
the charge carriers will demonstrate higher mobility along the $\Gamma\rightarrow Y$
direction in comparison to the $\Gamma\rightarrow X$ direction.

Moving from single-electron properties (i.e., electronic levels) to
two-particle properties, such as optical absorption, an additional
step in the many-body description is necessary. This involves accounting
for the Coulombic interactions between excited electrons in the conduction
band and holes generated in the valence band, resulting in the formation
of a bound pair of electron and hole, i.e., an exciton. These excitonic
effects are crucial for understanding the optical properties of 2D-materials
\citep{PhysRevLett.104.226804,PhysRevLett.103.186802}. To incorporate
these excitonic effects, the Bethe-Salpeter equation (BSE) is solved
on top of G$_{0}$W$_{0}$ corrected energies. While performing G$_{0}$W$_{0}$
calculations, we selected orbitals from HSE06 as the starting point,
a critical factor for accurately estimating optical properties. 

\begin{figure}[h]
\includegraphics[scale=0.6]{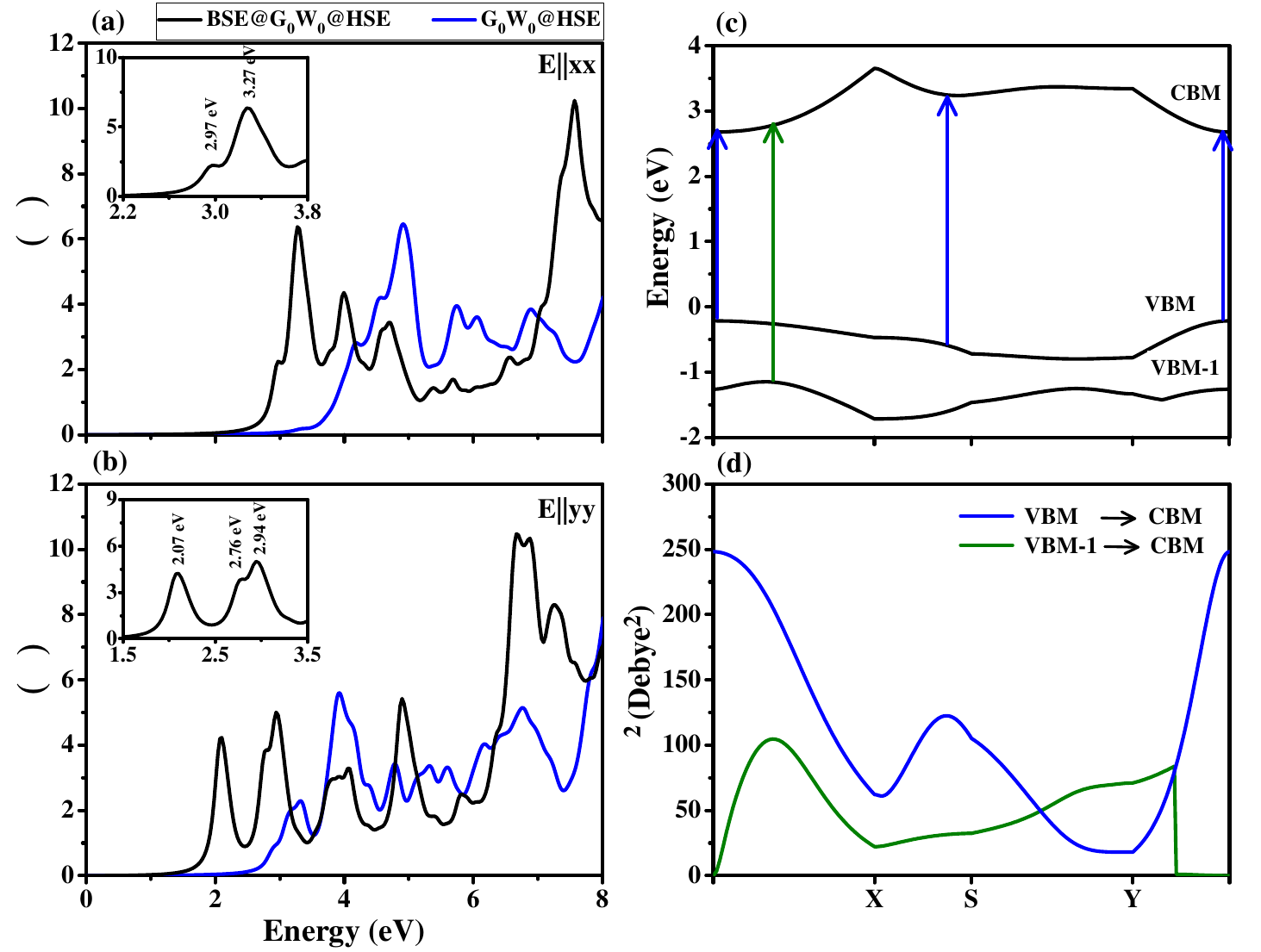}

\caption{\label{fig:figure-03}(a) \& (b) correspond to the excitonic absorption
spectrum for light polarized along the x-direction (E||xx) and y-direction
(E||yy), obtained using different levels of theory: G$_{0}$W$_{0}$@HSE
(blue solid line) and BSE@G$_{0}$W$_{0}$@HSE (black solid line).
c) Electronic band structure, and d) Transition dipole moment (TDM)
for the 2D SiC-biphenylene monolayer. The allowed dipole transitions
are shown with colored arrows.}
\end{figure}

Due to the significant depolarization effect in 2D materials for light
polarization perpendicular to the surface, the optical spectrum for
E||z polarization is negligible. Consequently, our analysis will concentrate
exclusively on the in-plane polarizations (i.e., E||xx and E||yy).
In Fig. (\ref{fig:figure-03}a-b), we compare the optical absorption
spectra of monolayer SiC-biphenylene computed at two levels of theory:
one with the inclusion of electron-hole interaction (BSE@G$_{0}$W$_{0}$@HSE),
and the other without electron-hole interaction (G$_{0}$W$_{0}$@HSE).
The introduction of electron-hole interaction results in a red-shift
in the optical spectra with strongly bound exciton states below the
G$_{0}$W$_{0}$ band gap (2.89 eV), which are notably absent in the
G$_{0}$W$_{0}$@HSE spectrum. As illustrated in Fig. (\ref{fig:figure-03}a-b),
the absorption spectra for light polarization along the x and y axes
(i.e., E||xx and E||yy) differ significantly, with the absorption
edge blue-shifted for E||xx. This strong effect of light polarization
on the absorption spectra is primarily due to the anisotropic structure
of the material along the x and y directions. In anisotropic materials,
the structural arrangement, such as the orientation of atoms and bonds,
varies along different crystallographic axes. This structural anisotropy
influences how the material interacts with polarized light, as the
interaction depends on the alignment of the material's structure relative
to the polarization direction of the light. Consequently, light polarized
along the x or y axis interacts differently with the material, resulting
in distinct absorption characteristics at specific energies. 

We identified two optically active (bright) excitons at 2.07 eV and
2.76 eV, both of which lie below the $G_{0}W_{0}$ band gap of 2.89
eV and are formed by light polarized in the y-direction. The calculated
binding energy of the first exciton, located at 2.07 eV, is determined
to be 0.82 eV. Although this value is lower than the theoretically
predicted binding energies for h-SiC (1.11 eV) \citep{PhysRevB.84.085404}
and Si$_{9}$C$_{15}$ (1.14 eV) \citep{PhysRevB.107.085114}, it
remains significantly higher than the predicted exciton binding energy
for bulk 2H-SiC, which is 0.1 eV \citep{PhysRevB.84.085404}. These
findings suggest that excitons in 2D materials experience strong confinement
due to reduced dimensionality, which enhances the overlap of electron
and hole wave functions and results in pronounced excitonic effects.
The calculated Bohr radius and effective mass of the first exciton,
generated by y-polarized light, are 2.14 $\text{Å}$ and 1.01 $m_{0}$,
respectively. Notably, the Bohr radius is smaller than the lattice
parameters ($|\vec{a}|$ = 5.61 $\text{Å}$, $|\vec{b}|$= 9.44 $\text{Å}$),
further indicating the strong confinement of the exciton within the
unit cell. This is a characteristic of the Frenkel exciton \citep{PhysRevMaterials.6.014012}.
In addition to the bright excitons generated by y-polarized light,
we identified the first bright exciton generated by x-polarized light
at 2.97 eV. This highlights the anisotropic nature of the optical
response of excitons within the 2D SiC-biphenylene monolayer, driven
by the interplay between the reduced dimensionality and the inherent
anisotropy in the crystal lattice. Apart from optically active (bright)
excitons, there are many optically inactive (dark) excitons. However,
we find no dark excitons below the first bright exciton, corroborating
the “direct” band gap nature of monolayer SiC-biphenylene. Notably,
a cluster of three dark excitons with nearly identical energies appears
just below the second bright exciton at 2.76 eV.

Next, we calculated the transition probabilities from the valence
band maximum (VBM) to the conduction band minimum (CBM), as shown
in Fig. (\ref{fig:figure-03}c-d). The highest magnitude of the transition
dipole moment (TDM) for the VBM to CBM transition occurs at the $\Gamma$-point.
In contrast, the TDM is minimum at the X and Y points for this transition,
indicating that optical absorption is forbidden at these points. For
the transition between VBM-1 and CBM, the transition probability is
highest between the $\Gamma$ and X points in the Brillouin zone.

\subsection{Elastic properties \label{subsec:c}}

\begin{figure}[h]
\includegraphics[clip,scale=0.6]{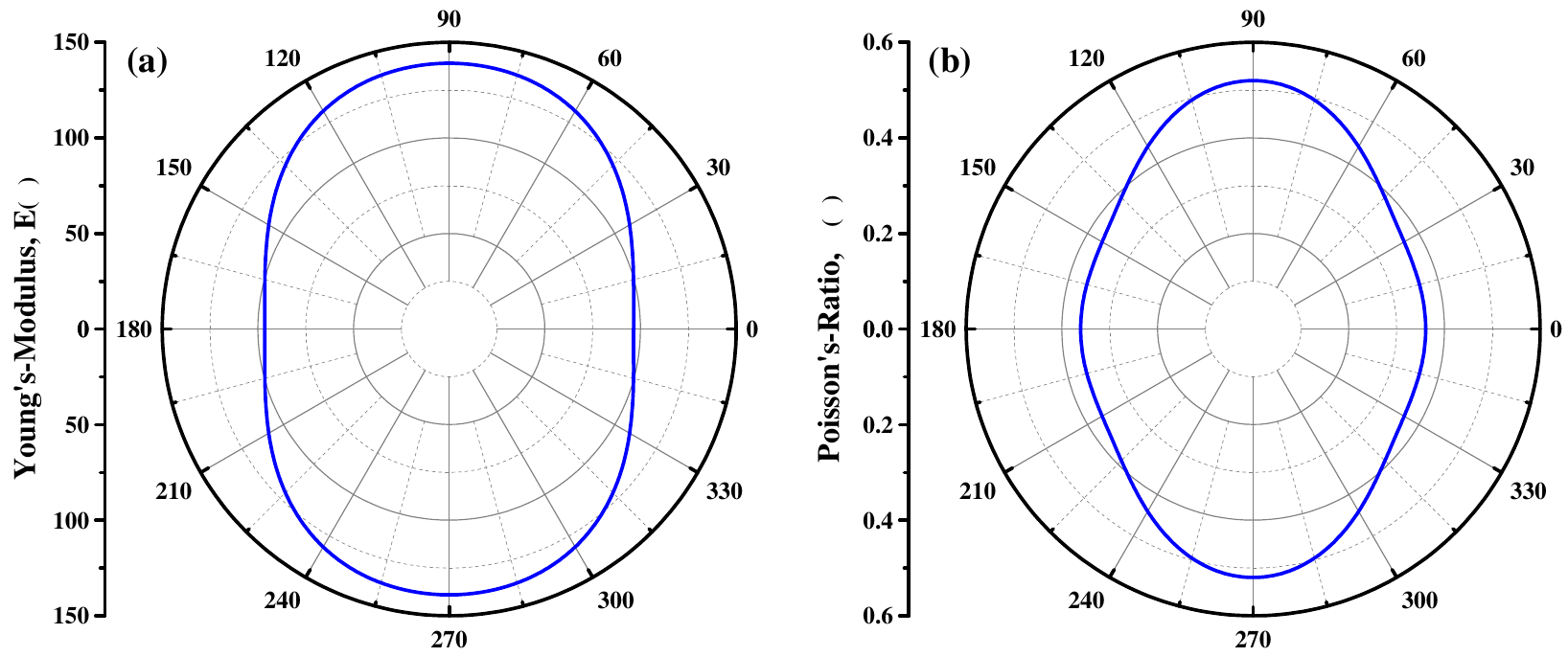}\caption{\label{fig:figure-04}Calculated orientation dependent (a) Young's
Modulus $E(\theta)$, and (b) Poisson's Ratio $\nu(\theta)$ of a
2D SiC-biphenylene monolayer}

\end{figure}

Gaining insights into the mechanical properties of a material is essential
for its effective utilization in device manufacturing. Strain engineering
is one of the prevalent strategies for tailoring the properties of
nanomaterials. Strains can occur due to the mismatch of lattice constants
with the substrate. Hence, a comprehensive knowledge of mechanical
properties is highly desirable. The full reversible lattice response
to the small strain around equilibrium is quantified by the elastic
modulus. To determine the elastic constants, we applied a series of
strains ranging from -2\% to +2\% with a step size of 0.05\%, assessing
the total energy of the material at each interval. Subsequently, by
fitting the energy-strain curves using Eqn.\ref{eq:eqn-6}, we obtained
the elastic constants: $C_{11}$ = 118.66, $C_{12}$ = 61.66, $C_{22}$
= 171.00, and $C_{66}$ = 43.56, respectively. The elastic constants
clearly satisfy the Born--Huang criteria \citep{born1955dynamical}
$C_{11}C_{22}-C_{12}^{2}>0$ and $C_{66}>0$, indicating the mechanical
stability of the material. To estimate the mechanical strength, the
direction-dependent Young's Modulus $E(\theta)$ and Poisson's Ratio
$\nu(\theta)$ are calculated based on the calculated elastic constants
using Eqns.\ref{eq:eqn-7} \& \ref{eq:eqn-8}. The calculated Young's
Modulus $E(\theta)$ of SiC-biphenylene along x $(\sim0\lyxmathsym{\textdegree})$
and y $(\sim90\lyxmathsym{\textdegree})$ directions is 96.44 N/m
and 138.97 N/m, respectively, implying the mechanical anisotropy of
the system, as shown in Fig. (\ref{fig:figure-04}a). The values are
notably lower than those of biphenylene and graphene structures (Table
\ref{tab:tab-2}), likely due to the larger lattice constant and lower
symmetry of SiC-biphenylene relative to Biphenylene and graphene.
The correponding Poisson's ratio along x $(\sim0\lyxmathsym{\textdegree})$
and y $(\sim90\lyxmathsym{\textdegree})$ directions are 0.36 and
0.52, respectively (\ref{fig:figure-04}b). The positive Poisson's
ratio indicates that when a compressive (tensile) strain is applied
in one direction, SiC-biphenylene tends to expand (contract) in the
perpendicular direction \citep{Greaves2011}. The relatively low Young's
modulus of SiC-biphenylene offers the potential to utilize it in flexible
electronics and to tailor its mechanical properties through strain
engineering.

\begin{table}[h]
\begin{tabular}{>{\centering}p{3.5cm}>{\centering}p{3.5cm}>{\centering}p{3.5cm}>{\centering}p{3.5cm}>{\centering}p{3.5cm}}
\toprule 
\multirow{2}{3.5cm}{Material} & \multicolumn{2}{c}{Young's Modulus (N/m)} & \multicolumn{2}{c}{Poisson's Ratio}\tabularnewline
\cmidrule{2-5} \cmidrule{3-5} \cmidrule{4-5} \cmidrule{5-5} 
 & $E_{x}$ & $E_{y}$ & $\nu_{x}$ & $\nu_{y}$\tabularnewline
\midrule
\midrule 
SiC-biphenylene & 96.44  & 138.97  & 0.36  & 0.52\tabularnewline
\midrule 
Biphenylene \citep{luo2021first} & 212.4  & 259.7 & 0.31  & 0.38\tabularnewline
\midrule 
h-SiC \citep{PhysRevB.80.155453} & 166  & 166  & 0.29  & 0.29 \tabularnewline
\midrule 
Graphene \citep{PhysRevB.80.155453,PhysRevB.82.235414} & 335  & 335  & 0.16  & 0.16 \tabularnewline
\bottomrule
\end{tabular}\caption{\label{tab:tab-2} Calculated values of Young's Modulus ($E_{x}$,
$E_{y}$) and Poisson's Ratio ($\nu_{x},$$\nu_{y}$) for 2D SiC-biphenylene
monolayer. Corresponding values of 2D Biphenylene and h-SiC monolayers
have also been included for the sake of comaprison. }
\end{table}

\subsection{Melting Point\label{subsec:d}}

\begin{figure}[h]
\includegraphics[clip,scale=0.35]{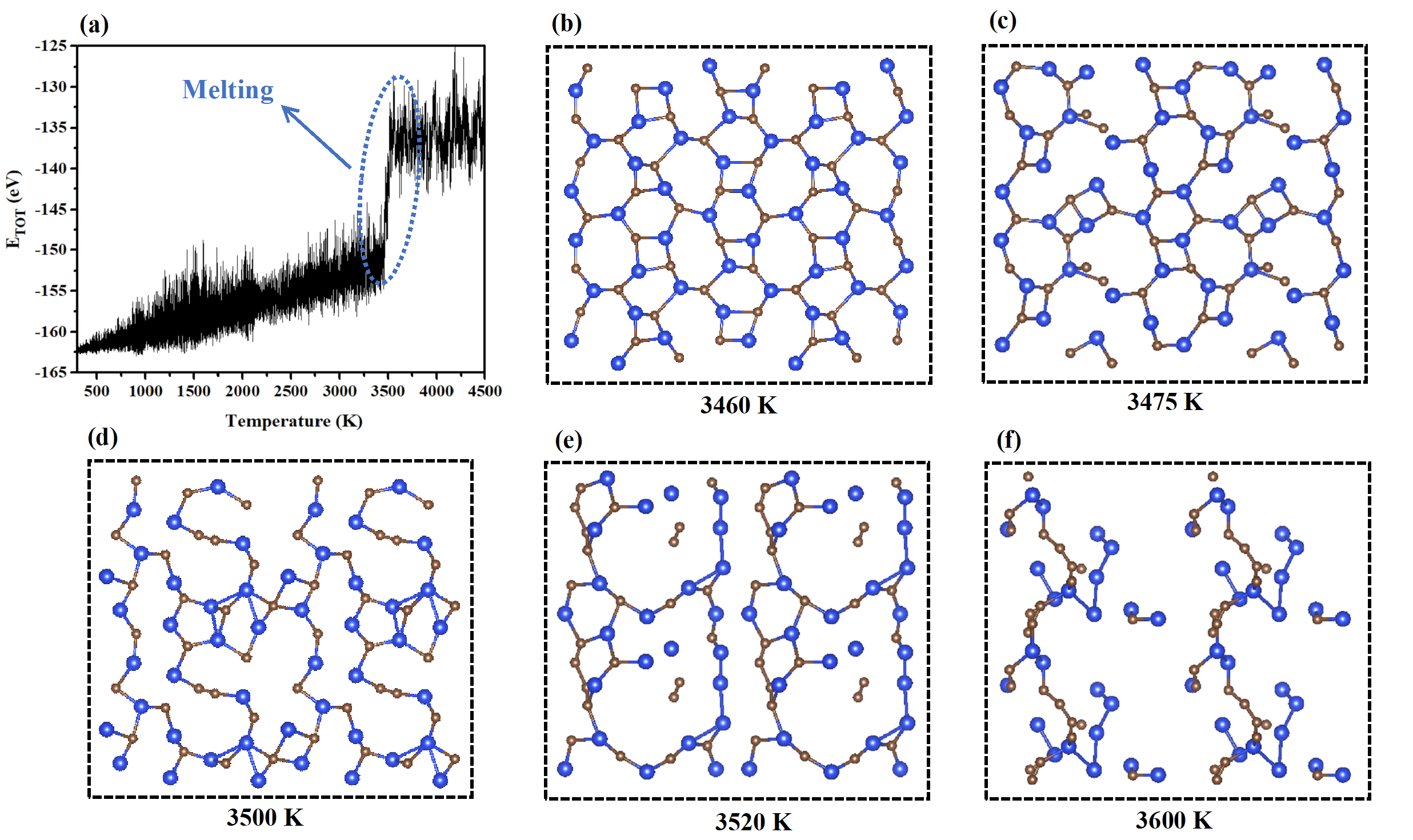}\caption{\label{fig:figure-05}(a) Variation of total energy E$_{TOT}$(eV)
with temperature ranging from 300 K to 4500 K for 2D SiC-biphenylene
monolayer. A suddent jump in E$_{TOT}$(eV) is seen between 3460 K
and 3520 K. MD snapshots of structure at (b) 3460 K , (c) 3475 K,
(d) 3500 K, (e) 3520 K, and (c) 3600 K. Bond breakage is observed
at 3475 K, signifying the onset of the melting process.}

\end{figure}

We conducted a comprehensive analysis of the thermal stability and
melting behavior of SiC-biphenylene using \emph{ab-intio} Molecular
Dynamics (MD) simulations across a temperature range varying from
300 K to 4500 K. Fig.\ref{fig:figure-05}a represents the variation
of total energy with respect to temperature. The total energy increases
linearly from 300 K to 3460 K, followed by an abrupt jump within the
range of 3460 K to 3520 K. This sudden surge in total energy is attributed
to increased atomic kinetic energy driven by the intense lattice thermal
vibrations, marking the onset of the melting process between 3460
K and 3520 K. During this temperature interval, we monitored corresponding
changes in structural morphology. At 3460 K, the structure retains
its integrity, experiencing bond deformation without any bond breakage.
However, the structure is highly strained due to robust thermal fluctuations
at the elevated temperature. Beyond 3475 K, the lattice thermal vibrations
intensify significantly, resulting in the rupture of Si-C bonds (Fig.\ref{fig:figure-05}c),
initiating the melting process. Consequently, SiC-biphenylene exhibits
a high melting point comparable to that of hexagonal-SiC (4050 K)
\citep{LeNguyen2020}, graphene (4095 K) \citep{Tromer2023} and biphenylene
(4024 K) \citep{D1NR07959J}. By 3500 K, the structure transitions
into entangled and interconnected chains of Si and C atoms, forming
a quasi-2D liquid state. Subsequently, at 3600 K (Fig.\ref{fig:figure-05}f),
clustering of Si and C atoms occurs, ultimately leading to the complete
dissociation of the material.

\subsection{Bilayer and bulk structures\label{subsec:e}}

\begin{figure}[h]
\includegraphics[clip,scale=0.9]{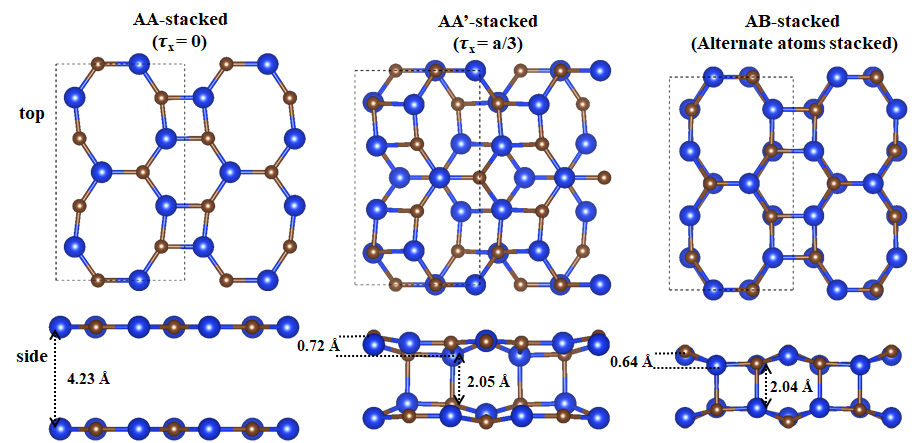}\caption{\label{fig:figure-06}Optimised atomic structure of SiC-biphenylene
bilayer with different stacking patterns : AA-stacked, AA'-stacked
and AB-stacked. The interlayer distance and buckling height have been
marked in the figure.}

\end{figure}

We constructed the bilayers of SiC-biphenylene by stacking the two
monolayers of SiC-biphenylene in various stacking configurations:
AA-stacked (where the top layer is aligned over the bottom layer with
zero translation, $\tau_{x}=0$); AA'-stacked (where the top layer
is offset from the bottom layer with a translation, $\tau_{x}=\frac{a}{3}$);
and AB-stacked (where the Si and C atoms are interchanged, i.e., the
Si atoms of the top layer align with the C atoms of the bottom layer,
and vice versa). These configurations are depicted in Fig. (\ref{fig:figure-06}).
To evaluate the interlayer binding energies in bilayer stackings,
we initially employed the semi-empirical DFT-D3 method to account
for van der Waals (vdW) interactions. For the AA-stacked bilayer,
the computed interlayer binding energy is -22.5 meV/atom, which is
comparable to that of the biphenylene bilayer (-22.0 meV/atom) \citep{PhysRevB.105.035408,PhysRevLett.115.115501}.
This suggests that the AA-stacked layers are predominantly stabilized
by weak van der Waals interactions, maintaining a planar geometry
with an interlayer distance of 4.23 Å. In contrast, the interlayer
binding energies for the AA'- and AB-stacked bilayers are significantly
higher, at -110.8 meV/atom and -173.7 meV/atom, respectively, suggesting
the possibility of weak vertical bond formation in addition to van
der Waals interactions. We further observe a buckling of 0.72 and
0.64 $\text{Å}$ in AA'- and AB-stacked geometries, respectively
(refer to Fig. \ref{fig:figure-06}). In order to capture vdW interactions
more effectively, we extended our analysis by employing non-local
vdW functionals, specifically vdW-opt-b86b and vdW-opt-b88. The results
reveal that the non-local functionals predict stronger binding energies
across all stackings compared to DFT-D3 (Table \ref{tab:table-04}),
reflecting their ability to capture long-range vdW interactions more
effectively. Notably, the non-local vdW potentials predict even stronger
binding in AA'- and AB-stacked configurations as compared to AA-stacked
configuration. Although the binding energies differ between methods,
the bond lengths remain nearly identical, suggesting that the equilibrium
geometries are largely unaffected by the choice of vdW corrections.
We observed that the buckling in the AA'-stacked bilayer structure
can be eliminated through the intercalation of noble gases such as
Ar, Kr, and Xe, resulting in interlayer distances of 6.63 $\text{Å}$,
6.92 $\text{Å}$, and 7.35 $\text{Å}$, respectively. The very recent
experimental investigations of the encapsulation of noble gas atoms
such as krypton (Kr), (Xe) and alkali metal atoms within bilayer graphene
have opened new avenues for research on encapsulated two-dimensional
van der Waals solids, with potential applications in quantum information
technology and energy storage \citep{langle2024two,lin2024alkali}. 

\begin{figure}[h]
\includegraphics[clip,scale=0.65]{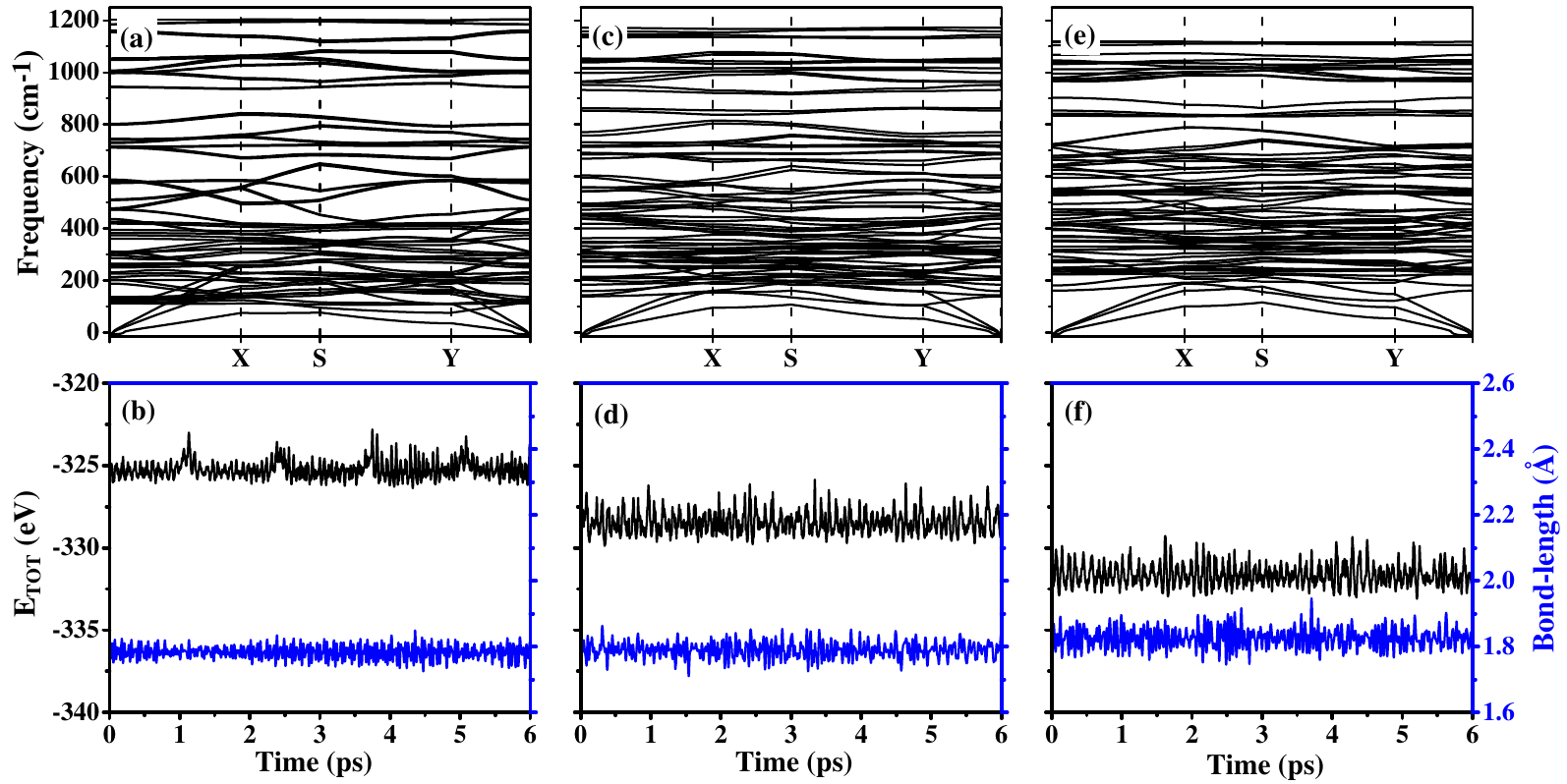}\caption{\label{figure-07}(a), (c), and (e) illustrate the calculated phonon
dispersion along high symmetry directions for AA, AA', and AB stacked
bilayers. In (b), (d), and (f), the variation in total energy and
bond lengths at 300 K is depicted for AA, AA', and AB stacked bilayers,
respectively. }
\end{figure}

Fig. (\ref{figure-07}a, c, and e) illustrates the phonon band dispersion
for AA, AA', and AB-stacked bilayer systems, respectively. The absence
of imaginary vibration frequencies across the Brillouin zone confirms
the dynamic stability of these stacking configurations. To confirm
their thermal stability at room temperature (300 K), AIMD simulations
were conducted, revealing that the structural integrity is preserved.
For the AA, AA', and AB-stacked bilayer systems, the total energy
fluctuations are approximately 2.42 eV ($\sim$0.74\%), 2.60 eV ($\sim$0.80\%),
and 2.37 eV ($\sim$0.72\%), respectively, with $\overline{Si_{2}C_{1}}$
bond length fluctuations of 0.07 $\text{Å}$, 0.12 $\text{Å}$,
and 0.12 $\text{Å}$ for each corresponding stacking configuration.
Fluctuations in other bond lengths are observed within a similar range,
from 0.06 $\text{Å}$ to 0.16 $\text{Å}$. Furthermore, to evaluate
their mechanical stability, the equilibrium systems were strained
by small amounts (-2\% to +2\%) to determine the elastic constants.
The derived elastic constants satisfy the Born--Huang criteria\citep{born1955dynamical}
$C_{11}C_{22}-C_{12}^{2}>0$ and $C_{66}>0$, confirming the thermal
stability of AA, AA', and AB-stacked bilayer systems.

\begin{figure}[h]
\includegraphics[clip,scale=0.65]{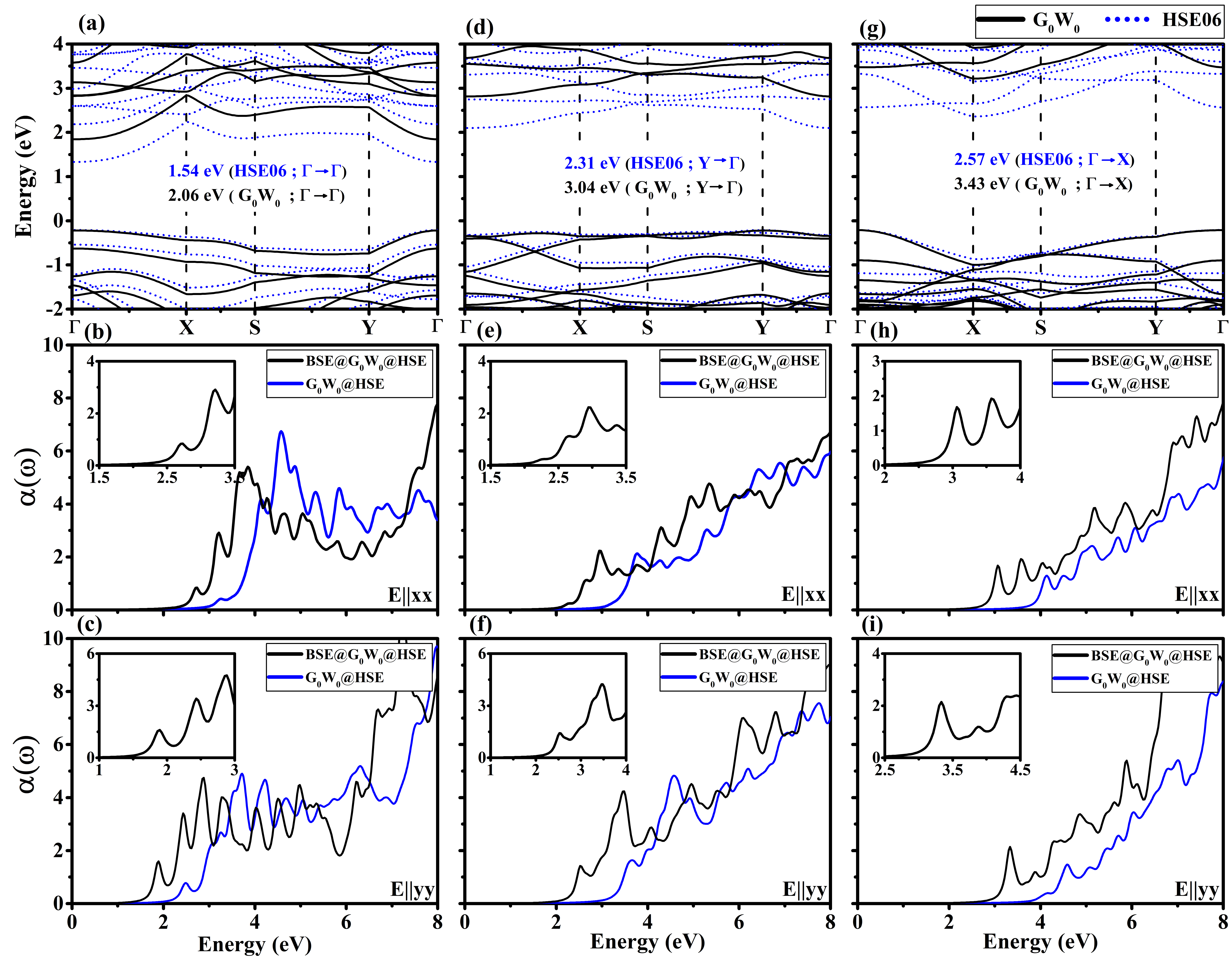}\caption{\label{fig:figure-08} (a), (d), and (g) show electronic band structures
for AA, AA', and AB stacked bilayers at HSE06 (blue dotted line) and
G$_{0}$W$_{0}$ (black solid line) levels of theory, with E$_{F}$
set at zero. (b) and (c) depict excitonic optical spectra for AA-stacked
structures. (e) and (f) represent excitonic optical spectra for AA'-stacked
structures. (h) and (i) illustrate excitonic optical spectra for AB-stacked
structures, calculated using G$_{0}$W$_{0}$@HSE (blue solid line)
and BSE@G$_{0}$W$_{0}$@HSE (black solid line)}
\end{figure}

To further investigate the influence of bilayer interactions across
various stacking configurations, we conducted band structure computations
employing HSE06 and G$_{0}$W$_{0}$ methodologies. The band gap and
quasi-particle (QP) self-energy corrections undergo significant variations
for different stacking patterns (\ref{tab:table-03}). It is evident
from Fig. (\ref{fig:figure-08}a) that the electronic characteristic
of AA-stacked bilayer resembles their constituent monolayers with
a relative shift in energy attributed to the weak van der Waals interactions.
The band gap narrows to 2.06 eV (G$_{0}$W$_{0}$); however, nature
remains ``direct'' at the $\Gamma$-point of the Brillouin zone.
In contrast, the electronic structure of AA'- and AB-stacked bilayers
significantly differs from their constituent monolayers due to the
vertical chemical interaction between the layers, in addition to van
der Waals interactions. This interaction results in reduction of interlayer
spacing and the formation of vertical bonds. Additionally, the atoms
within the bilayers arrange themselves into hexagonal and tetragonal
rings due to the formation of vertical bonds (Fig.\ref{fig:figure-06}).
The band gap transitions to an ``indirect'' nature and increases
to 3.04 eV (with the valence band maximum (VBM) at Y and the conduction
band minimum (CBM) at $\Gamma$) for AA'-stacked bilayers, and to
3.43 eV (with the VBM at $\Gamma$ and the CBM at X) for AB-stacked
bilayers, as shown in Fig. (\ref{fig:figure-08}d-g). It is evident
that for the bilayers, both the nature and the magnitude of the band
gap strongly depend on the stacking order.

\begin{table}[h]
\begin{tabular}{|>{\centering}p{2.1cm}|>{\centering}p{2cm}|>{\centering}p{2cm}|>{\centering}p{2.5cm}|>{\centering}p{2.5cm}|>{\centering}p{1.5cm}|>{\centering}p{1.5cm}|>{\centering}p{1.5cm}|>{\centering}p{1.5cm}|}
\hline 
SiC-biphenylene & $E_{g}$(HSE06)

(eV) & $E_{g}^{QP}$(G$_{0}$W$_{0}$)

(eV) & $m_{e}$

($m_{0}$) & $m_{h}$

($m_{0}$) & $E_{1}^{exciton}$

(eV) & $E_{b}^{exciton}$

(eV) & $\mu^{exciton}$

($m_{0}$) & $a^{exciton}$

($\text{Å})$\tabularnewline
\hline 
\hline 
monolayer & 2.16 

(Direct) & 2.89 

(Direct) & 1.34 $(\Gamma\rightarrow X)$ 0.25 $(\Gamma\rightarrow Y)$  & 2.21 $(\Gamma\rightarrow X)$ 0.28 $(\Gamma\rightarrow Y)$ & 2.07

(E||yy) & 0.82 & 1.01 & 2.14\tabularnewline
\hline 
AA-stacked

bilayer & 1.54 

(Direct) & 2.06 

(Direct) & 1.57 $(\Gamma\rightarrow X)$

0.21 $(\Gamma\rightarrow Y)$  & 2.90 $(\Gamma\rightarrow X)$

0.32 $(\Gamma\rightarrow Y)$ & 1.89

(E||yy) & 0.17 & 0.16 & 11.80\tabularnewline
\hline 
AA'-stacked

bilayer & 2.31 

(Indirect) & 3.04 

(Indirect) & 2.04 $(\Gamma\rightarrow X)$

0.34 $(\Gamma\rightarrow Y)$  & 0.95 $(Y\rightarrow S)$

0.87 $(Y\rightarrow\Gamma)$ & 2.52

(E||yy) & 0.64 & 0.55 & 3.28\tabularnewline
\hline 
AB-stacked

bilayer & 2.57

(Indirect) & 3.43 

(Indirect) & 0.05 $(X\rightarrow\Gamma)$

0.47 $(X\rightarrow S)$ & 0.44 $(\Gamma\rightarrow X)$

2.18 $(\Gamma\rightarrow Y)$ & 3.06

(E||xx) & 0.64 & 0.44 & 3.69\tabularnewline
\hline 
bulk (ABA...-stacked) & 2.78 

(Direct) & 3.01 

(Direct) & 1.30 $(\Gamma\rightarrow X)$

0.18 $(\Gamma\rightarrow Y)$ & 0.45 $(\Gamma\rightarrow X)$

0.89 $(\Gamma\rightarrow Y)$ & 3.21

(E||xx) & 0.20 & 0.31 & 7.88\tabularnewline
\hline 
\end{tabular}\caption{\label{tab:table-03} Band gaps calculated using HSE06 ($E_{g}$)
and G$_{0}$W$_{0}$($E_{g}^{QP}$). The electron ($m_{e}$) and hole
($m_{h}$) effective masses calculated along the direction of their
band extrema. First excitonic energy ($E_{1}^{exciton}$) and its
corresponding binding energy ($E_{b}^{exciton}$) , effective mass
($\mu^{exciton}$), and Bohr radius ($a^{exciton}$) are listed for
the monolayer, bilayer, and bulk systems.}
\end{table}

In Table \ref{tab:table-03}, we present the estimated effective masses
of electrons (holes) in the direction from CBM (VBM) to the nearest
high-symmetry point in the Brillouin Zone. Consistent with the band
structure calculations, the effective masses exhibit directional dependence
due to the anisotropy of the energy dispersion curves. Specifically,
in the AA-stacked configuration, the effective mass of electrons and
holes along the $\Gamma\rightarrow Y$ direction demonstrates an approximately
88\% reduction compared to that along the $\Gamma\rightarrow X$ direction.
Notably, the effective mass of electrons is 0.054 for $X\rightarrow\Gamma$
in the AB-stacked bilayer, suggesting that carriers behave more like
free particles with high mobility. This characteristic is advantageous
for semiconductor applications requiring fast charge transport and
efficient device performance.

To explore the influence of stacking patterns on excitonic effects,
we analyze the optical properties of bilayer configurations with varying
stacking arrangements. Notable distinctions emerge among the absorption
spectra across these varied configurations. The first bright excitonic
peak is observed at 1.89 eV (E||yy) for AA-stacked configurations,
2.52 (E||yy) eV for AA\textasciiacute -stacked, and 3.06 eV (E||xx)
for AB-stacked configurations, with the corresponding binding energies
of 0.17 eV, 0.64 eV, and 0.64 eV, respectively (Table \ref{tab:table-03}).
Distinct excitonic behaviors are evident across these stacking configurations,
with AA-stacked configurations exhibiting the presence of Mott-Wannier
excitons. In contrast, excitons in AA'-stacked and AB-stacked configurations
are significantly more localized compared to those in the AA-stacked
bilayer.

\begin{figure}[h]
\includegraphics[clip,scale=0.6]{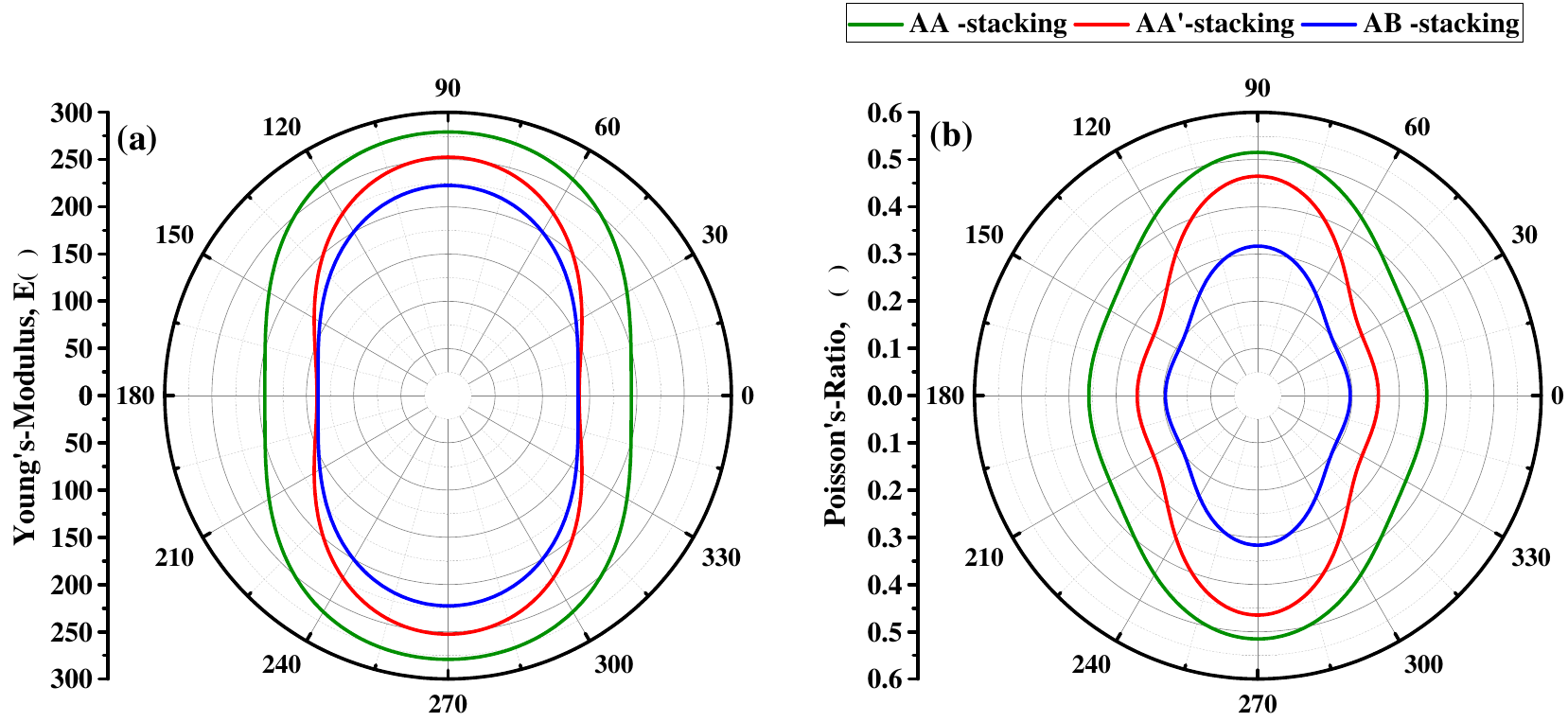}

\caption{\label{fig:figure-09}Calculated orientation dependent (a) Young's
Modulus $E(\theta)$ and (b) Poisson's Ratio $\nu(\theta)$ for AA,
AA', and AB stacked bilayers.}
\end{figure}

Furthermore, we examined the mechanical strength of AA, AA', and AB-stacked
bilayer systems by computing the direction-dependent Young's Modulus
and Poisson's ratio (Fig. \ref{fig:figure-09}). Young's Modulus values
increase (indicating higher stiffness) for bilayer stackings, with
the highest observed for AA-stacked bilayers. Clearly, the number
of layers and atomic stacking significantly affect the material's
stiffness. The minimum and maximum values of Young's Modulus are observed
along the x and y directions, respectively, similar to the monolayer
case of SiC-biphenylene (Table \ref{tab:table-04}). The Poisson's
ratio for AA-stacked bilayers remains the same as that obtained for
monolayers, which can be attributed to weak van der Waals interactions
between the layers. However, for AA' and AB-stacked bilayers, weak
vertical bonds are formed, which largely reduce the property of lateral
contraction upon extension, resulting in lower Poisson's values compared
to monolayer SiC-biphenylene.

\begin{table}[h]
\begin{tabular}{>{\centering}p{2.5cm}>{\centering}p{2cm}>{\centering}p{2cm}>{\centering}p{2cm}>{\centering}p{2cm}>{\centering}m{2cm}>{\centering}m{2cm}>{\centering}m{2cm}}
\toprule 
\multirow{2}{2.5cm}{Stacking-type} & \multicolumn{2}{c}{Young's Modulus (N/m)} & \multicolumn{2}{c}{Poisson's Ratio} & \multicolumn{3}{>{\centering}p{4cm}}{Interlayer binding energy (meV/atom)}\tabularnewline
\cmidrule{2-8} \cmidrule{3-8} \cmidrule{4-8} \cmidrule{5-8} \cmidrule{6-8} \cmidrule{7-8} \cmidrule{8-8} 
 & $E_{x}$ & $E_{y}$ & $\nu_{x}$ & $\nu_{y}$ & DFT-D3 & optB86b-vdW  & optB88-vdW\tabularnewline
\midrule 
AA-stacked & 193.90 & 279.15 & 0.36 & 0.51 & -22.5 & -32.08 & -32.92\tabularnewline
\midrule 
AA'-stacked & 138.77 & 252.49 & 0.25 & 0.46 & -110.83 & -142.08 & -121.25\tabularnewline
\midrule 
AB-stacked & 137.56 & 222.52 & 0.19 & 0.32 & -173.75 & -208.75 & -171.67\tabularnewline
\bottomrule
\end{tabular}\caption{\label{tab:table-04}Calculated values of Young's Modulus ($E_{x}$,
$E_{y}$) and Poisson's Ratio ($\nu_{x},$$\nu_{y}$) for different
bilayer stackings. The interlayer binding energies were also computed
using semi-empirical DFT-D3 and non-local vdW functionals (vdW-opt-b86b
and vdW-opt-b88). }
\end{table}

\begin{figure}[h]
\includegraphics[clip,scale=0.55]{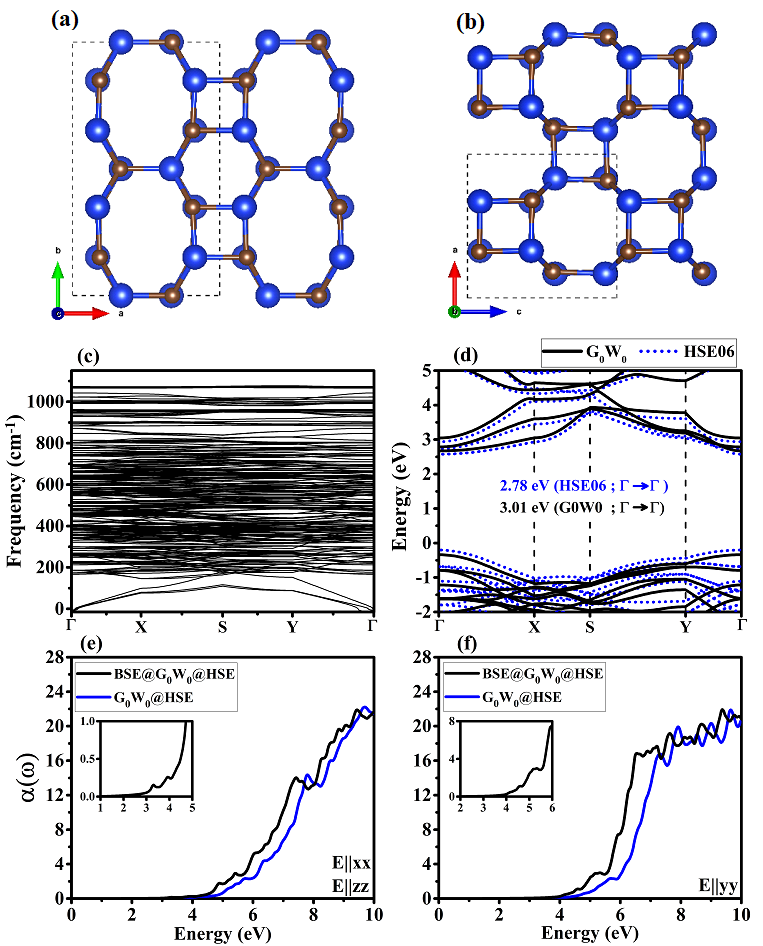}

\caption{\label{fig:figure-10} Optimised structure of bulk-SiC-biphenylene
(stacked in ABAB... configuration) projected on (a) xy-plane, (b)
xz-plane, (c) calculated phonon dispersion, (d) electronic band structure
along the high symmetry directions of Brillouin zone, and (e) \& (f)
optical absorption spectrum calculated using G$_{0}$W$_{0}$@HSE
(blue solid line) and BSE@G$_{0}$W$_{0}$@HSE (black solid line)}
\end{figure}

Among the three bilayer stacking geometries, the AB geometry exhibits
the highest binding energy, estimated at around -173.7 meV/atom, indicating
superior stability for the formation of bulk structures. Consequently,
we opt for an AB stacking configuration when constructing the bulk
structure of SiC-biphenylene. The absence of imaginary frequencies
(Fig.\ref{fig:figure-10}c) confirms dynamical stability at 0 K. Furthermore,
thermal stability analysis conducted through AMID calculations demonstrates
that bulk-SiC-biphenylene maintains its stability at room temperature.
The vertical chemical bonds are reinforced, providing a robust interlayer
binding energy of -395 meV/atom. The structure exhibits isotropy along
the xy and yz planes, featuring rings of octagons, hexagons, and tetragons
(Fig.\ref{fig:figure-10}a). In contrast, along the xz-plane, only
octagon and tetragon rings are observed (Fig.\ref{fig:figure-10}b).
The quasi-particle band gap is calculated to be 3.01 eV, with a self-energy
correction of 0.23 eV to the HSE06 value (Fig.\ref{fig:figure-10}d).
The lower magnitude of the self-energy correction compared to monolayer
and bilayer systems can be attributed to the dimensionality effect,
where electrons are more delocalized in the bulk, thus reducing the
impact of electron-electron correlation. Bulk SiC-biphenylene exhibits
its first excitonic peak at 3.21 eV (E||xx) with a binding energy
of 0.20 eV (Fig.\ref{fig:figure-10}e). The corresponding Bohr radius
of the exciton is 7.88 Å, indicating that this peak corresponds to
a Mott-Wannier exciton. As the material transitions from monolayer
to bulk, excitonic optical absorptions become less significant.

\section{CONCLUSION\label{sec:conclusion}}

In this article, we have computationally designed a stable novel SiC
allotrope within a biphenylene network, comprising interconnected
polygons of octagons, hexagons, and tetragons arranged in a periodic
fashion. The structure exhibits excellent thermal stability, with
a high melting point of 3475 K. Computational analysis of its mechanical
properties indicates that SiC-biphenylene is softer compared to graphene
and biphenylene, making it suitable for high-power flexible electronic
devices. Our findings indicate that it is a ``direct'' band gap
semiconductor with a quasiparticle (QP) band gap value of 2.89 eV
(G$_{0}$W$_{0}$). The atom-projected Density of States (DOS) reveals
that the Conduction Band Minimum (CBM) and Valence Band Maximum (VBM)
are composed of the p-orbitals of Si and C atoms, respectively, which
may facilitate exciton separation in photovoltaic applications. Moreover,
the effective mass of charge carriers along the $\Gamma\rightarrow Y$
direction is found to be less than that along the $\Gamma\rightarrow X$
direction, suggesting enhanced mobility along the former path. Additionally,
a sharp excitonic peak below the QP band gap indicates the presence
of a strongly bound bright exciton with a binding energy of 0.82 eV.
Furthermore, the anisotropic optical excitations over a wide spectral
range result in polarization-dependent optical properties.

Our investigation extended to include bilayer structures formed in
three distinct stacking geometries (AA, AA', and AB-stacked), analyzing
their energetics, stability, as well as electronic and optical properties.
We found that both the nature and magnitude of band gap depends on
stacking order of bilayer structure. Additionally, we explored the
influence of stacking patterns on excitonic effects and observed notable
distinctions in the absorption spectra. Specifically, excitons are
found to be more localized in AA' and AB-stacked configurations compared
to the AA-stacked configuration. Furthermore, by examining specific
stacking configurations, we have identified the bulk structure of
SiC-biphenylene (ABAB...-stacked) and confirmed its characteristic
``direct'' band gap. As the material transitions from monolayer
to bulk, excitonic optical absorptions become less significant. In
summary, our study offers in-depth insights into the electronic and
optical characteristics of monolayer SiC-biphenylene, as well as its
bilayer and bulk configurations. We anticipate that the results presented
in this paper will broaden the understanding of the properties of
2D SiC nanomaterial allotropes.

\newpage{}

\bibliographystyle{plain}
\addcontentsline{toc}{section}{\refname}\bibliography{ref}

\begin{thebibliography}{10}

\bibitem{PhysRevB.85.125428}
R.~C. Andrew, R.~E. Mapasha, A.~M. Ukpong, and N.~Chetty.
\newblock Mechanical properties of graphene and boronitrene.
\newblock {\em Phys. Rev. B}, 85:125428, Mar 2012.

\bibitem{C5CP00601E}
C.~Attaccalite, A.~Nguer, E.~Cannuccia, and M.~Grüning.
\newblock Strong second harmonic generation in sic{,} zno{,} gan
  two-dimensional hexagonal crystals from first-principles many-body
  calculations.
\newblock {\em Phys. Chem. Chem. Phys.}, 17:9533--9540, 2015.

\bibitem{bafekry2021biphenylene}
A~Bafekry, M~Faraji, MM~Fadlallah, HR~Jappor, S~Karbasizadeh, M~Ghergherehchi,
  and D~Gogova.
\newblock Biphenylene monolayer as a two-dimensional nonbenzenoid carbon
  allotrope: a first-principles study.
\newblock {\em Journal of Physics: Condensed Matter}, 34(1):015001, 2021.

\bibitem{bandyopadhyay2020review}
Arka Bandyopadhyay and Debnarayan Jana.
\newblock A review on role of tetra-rings in graphene systems and their
  possible applications.
\newblock {\em Reports on Progress in Physics}, 83(5):056501, 2020.

\bibitem{Barone2011}
Veronica Barone, Oded Hod, Juan~E. Peralta, and Gustavo~E. Scuseria.
\newblock Accurate prediction of the electronic properties of low-dimensional
  graphene derivatives using a screened hybrid density functional.
\newblock {\em Accounts of Chemical Research}, 44(4):269--279, Apr 2011.

\bibitem{PhysRevB.81.075433}
E.~Bekaroglu, M.~Topsakal, S.~Cahangirov, and S.~Ciraci.
\newblock First-principles study of defects and adatoms in silicon carbide
  honeycomb structures.
\newblock {\em Phys. Rev. B}, 81:075433, Feb 2010.

\bibitem{PhysRevB.50.17953}
P.~E. Bl\"ochl.
\newblock Projector augmented-wave method.
\newblock {\em Phys. Rev. B}, 50:17953--17979, Dec 1994.

\bibitem{born1955dynamical}
Max Born, Kun Huang, and M~Lax.
\newblock Dynamical theory of crystal lattices.
\newblock {\em American Journal of Physics}, 23(7):474--474, 1955.

\bibitem{PhysRevB.82.235414}
Emiliano Cadelano, Pier~Luca Palla, Stefano Giordano, and Luciano Colombo.
\newblock Elastic properties of hydrogenated graphene.
\newblock {\em Phys. Rev. B}, 82:235414, Dec 2010.

\bibitem{D2CP04752G}
Xin-Wei Chen, Zheng-Zhe Lin, and Xi-Mei Li.
\newblock Biphenylene network as sodium ion battery anode material.
\newblock {\em Phys. Chem. Chem. Phys.}, 25:4340--4348, 2023.

\bibitem{crespi1996prediction}
Vincent~H Crespi, Lorin~X Benedict, Marvin~L Cohen, and Steven~G Louie.
\newblock Prediction of a pure-carbon planar covalent metal.
\newblock {\em Physical Review B}, 53(20):R13303, 1996.

\bibitem{PhysRevLett.104.226804}
Pierluigi Cudazzo, Claudio Attaccalite, Ilya~V. Tokatly, and Angel Rubio.
\newblock Strong charge-transfer excitonic effects and the bose-einstein
  exciton condensate in graphane.
\newblock {\em Phys. Rev. Lett.}, 104:226804, Jun 2010.

\bibitem{PhysRevB.105.035408}
Salih Demirci, \ifmmode \mbox{\c{S}}\else~\c{S}\fi{}afak \ifmmode
  \mbox{\c{C}}\else \c{C}\fi{}all\ifmmode \imath \else \i
  \fi{}o\ifmmode~\breve{g}\else \u{g}\fi{}lu, Taylan G\"orkan, Ethem Akt\"urk,
  and Salim Ciraci.
\newblock Stability and electronic properties of monolayer and multilayer
  structures of group-iv elements and compounds of complementary groups in
  biphenylene network.
\newblock {\em Phys. Rev. B}, 105:035408, Jan 2022.

\bibitem{fan2021biphenylene}
Qitang Fan, Linghao Yan, Matthias~W Tripp, Ond{\v{r}}ej Krej{\v{c}}{\'\i},
  Stavrina Dimosthenous, Stefan~R Kachel, Mengyi Chen, Adam~S Foster, Ulrich
  Koert, Peter Liljeroth, et~al.
\newblock Biphenylene network: A nonbenzenoid carbon allotrope.
\newblock {\em Science}, 372(6544):852--856, 2021.

\bibitem{ferguson2017biphenylene}
David Ferguson, Debra~J Searles, and Marlies Hankel.
\newblock Biphenylene and phagraphene as lithium ion battery anode materials.
\newblock {\em ACS applied materials \& interfaces}, 9(24):20577--20584, 2017.

\bibitem{adma.202204779}
Zhao-Yan Gao, Wenpeng Xu, Yixuan Gao, Roger Guzman, Hui Guo, Xueyan Wang,
  Qi~Zheng, Zhili Zhu, Yu-Yang Zhang, Xiao Lin, Qing Huan, Geng Li, Lizhi
  Zhang, Wu~Zhou, and Hong-Jun Gao.
\newblock Experimental realization of atomic monolayer si9c15.
\newblock {\em Advanced Materials}, 34(35):2204779, 2022.

\bibitem{ge2021superconductivity}
Yanfeng Ge, Zhicui Wang, Xing Wang, Wenhui Wan, and Yong Liu.
\newblock Superconductivity in the two-dimensional nonbenzenoid biphenylene
  sheet with dirac cone.
\newblock {\em 2D Materials}, 9(1):015035, 2021.

\bibitem{PhysRevLett.62.1169}
R.~W. Godby and R.~J. Needs.
\newblock Metal-insulator transition in kohn-sham theory and quasiparticle
  theory.
\newblock {\em Phys. Rev. Lett.}, 62:1169--1172, Mar 1989.

\bibitem{GODET2001168}
C.~Godet and M.N. Berberan-Santos.
\newblock Evidence for excitonic behavior of photoluminescence in polymer-like
  a-c:h films.
\newblock {\em Diamond and Related Materials}, 10(2):168--173, 2001.
\newblock Proceedings of the 3rd Specialist Meeting on Amorphous Carbon.

\bibitem{gorkan2023can}
Taylan Gorkan, Salih Demirci, Johannes~V. Barth, Ethem Akt{\"u}rk, and Salim
  Ciraci.
\newblock Can stable mos2 monolayers and multilayers be constituted in the
  biphenylene network?
\newblock {\em The Journal of Physical Chemistry C}, 127(18):8770--8777, May
  2023.

\bibitem{Greaves2011}
G.~N. Greaves, A.~L. Greer, R.~S. Lakes, and T.~Rouxel.
\newblock Poisson's ratio and modern materials.
\newblock {\em Nature Materials}, 10(11):823--837, Nov 2011.

\bibitem{10.1063/1.3382344}
Stefan Grimme, Jens Antony, Stephan Ehrlich, and Helge Krieg.
\newblock {A consistent and accurate ab initio parametrization of density
  functional dispersion correction (DFT-D) for the 94 elements H-Pu}.
\newblock {\em The Journal of Chemical Physics}, 132(15):154104, 04 2010.

\bibitem{https://doi.org/10.1002/jcc.21759}
Stefan Grimme, Stephan Ehrlich, and Lars Goerigk.
\newblock Effect of the damping function in dispersion corrected density
  functional theory.
\newblock {\em Journal of Computational Chemistry}, 32(7):1456--1465, 2011.

\bibitem{D2CP00798C}
Ting Han, Yu~Liu, Xiaodong Lv, and Fengyu Li.
\newblock Biphenylene monolayer: a novel nonbenzenoid carbon allotrope with
  potential application as an anode material for high-performance sodium-ion
  batteries.
\newblock {\em Phys. Chem. Chem. Phys.}, 24:10712--10716, 2022.

\bibitem{PhysRevB.102.134103}
Q.~Hassanzada, I.~Abdolhosseini Sarsari, A.~Hashemi, A.~Ghojavand, A.~Gali, and
  M.~Abdi.
\newblock Theoretical study of quantum emitters in two-dimensional silicon
  carbide monolayers.
\newblock {\em Phys. Rev. B}, 102:134103, Oct 2020.

\bibitem{10.1063/1.1760074}
Jochen Heyd and Gustavo~E. Scuseria.
\newblock {Efficient hybrid density functional calculations in solids:
  Assessment of the Heyd–Scuseria–Ernzerhof screened Coulomb hybrid
  functional}.
\newblock {\em The Journal of Chemical Physics}, 121(3):1187--1192, 07 2004.

\bibitem{10.1063/1.1564060}
Jochen Heyd, Gustavo~E. Scuseria, and Matthias Ernzerhof.
\newblock {Hybrid functionals based on a screened Coulomb potential}.
\newblock {\em The Journal of Chemical Physics}, 118(18):8207--8215, 05 2003.

\bibitem{10.1063/1.2204597}
Jochen Heyd, Gustavo~E. Scuseria, and Matthias Ernzerhof.
\newblock {Erratum: “Hybrid functionals based on a screened Coulomb
  potential” [J. Chem. Phys. 118, 8207 (2003)]}.
\newblock {\em The Journal of Chemical Physics}, 124(21):219906, 06 2006.

\bibitem{SACADA}
Roald Hoffmann, Artyom~A. Kabanov, Andrey~A. Golov, and Davide~M. Proserpio.
\newblock Homo citans and carbon allotropes: For an ethics of citation.
\newblock {\em Angewandte Chemie International Edition}, 55(37):10962--10976,
  2016.

\bibitem{PhysRevB.84.085404}
H.~C. Hsueh, G.~Y. Guo, and Steven~G. Louie.
\newblock Excitonic effects in the optical properties of a sic sheet and
  nanotubes.
\newblock {\em Phys. Rev. B}, 84:085404, Aug 2011.

\bibitem{Hudspeth}
Mathew~A. Hudspeth, Brandon~W. Whitman, Veronica Barone, and Juan~E. Peralta.
\newblock Electronic properties of the biphenylene sheet and its
  one-dimensional derivatives.
\newblock {\em ACS Nano}, 4(8):4565--4570, Aug 2010.

\bibitem{PhysRevB.80.155453}
H.~\ifmmode~\mbox{\c{S}}\else \c{S}\fi{}ahin, S.~Cahangirov, M.~Topsakal,
  E.~Bekaroglu, E.~Akturk, R.~T. Senger, and S.~Ciraci.
\newblock Monolayer honeycomb structures of group-iv elements and iii-v binary
  compounds: First-principles calculations.
\newblock {\em Phys. Rev. B}, 80:155453, Oct 2009.

\bibitem{PhysRevB.83.195131}
Ji\ifmmode \check{r}\else~\v{r}\fi{}\'{\i} Klime\ifmmode~\check{s}\else
  \v{s}\fi{}, David~R. Bowler, and Angelos Michaelides.
\newblock Van der waals density functionals applied to solids.
\newblock {\em Phys. Rev. B}, 83:195131, May 2011.

\bibitem{Klime_2010}
Jiří Klimeš, David~R Bowler, and Angelos Michaelides.
\newblock Chemical accuracy for the van der waals density functional.
\newblock {\em Journal of Physics: Condensed Matter}, 22(2):022201, dec 2009.

\bibitem{PhysRevB.54.11169}
G.~Kresse and J.~Furthm\"uller.
\newblock Efficient iterative schemes for ab initio total-energy calculations
  using a plane-wave basis set.
\newblock {\em Phys. Rev. B}, 54:11169--11186, Oct 1996.

\bibitem{KRESSE199615}
G.~Kresse and J.~Furthmüller.
\newblock Efficiency of ab-initio total energy calculations for metals and
  semiconductors using a plane-wave basis set.
\newblock {\em Computational Materials Science}, 6(1):15--50, 1996.

\bibitem{PhysRevB.59.1758}
G.~Kresse and D.~Joubert.
\newblock From ultrasoft pseudopotentials to the projector augmented-wave
  method.
\newblock {\em Phys. Rev. B}, 59:1758--1775, Jan 1999.

\bibitem{Kroto1985}
H.~W. Kroto, J.~R. Heath, S.~C. O'Brien, R.~F. Curl, and R.~E. Smalley.
\newblock C60: Buckminsterfullerene.
\newblock {\em Nature}, 318(6042):162--163, Nov 1985.

\bibitem{10.1063/1.2404663}
Aliaksandr~V. Krukau, Oleg~A. Vydrov, Artur~F. Izmaylov, and Gustavo~E.
  Scuseria.
\newblock {Influence of the exchange screening parameter on the performance of
  screened hybrid functionals}.
\newblock {\em The Journal of Chemical Physics}, 125(22):224106, 12 2006.

\bibitem{RevModPhys.80.3}
Stephan K\"ummel and Leeor Kronik.
\newblock Orbital-dependent density functionals: Theory and applications.
\newblock {\em Rev. Mod. Phys.}, 80:3--60, Jan 2008.

\bibitem{langle2024two}
Manuel L{\"a}ngle, Kenichiro Mizohata, Clemens Mangler, Alberto Trentino, Kimmo
  Mustonen, E~Harriet {\AA}hlgren, and Jani Kotakoski.
\newblock Two-dimensional few-atom noble gas clusters in a graphene sandwich.
\newblock {\em Nature Materials}, pages 1--6, 2024.

\bibitem{PhysRevB.78.205108}
H.~M. Lawler, J.~J. Rehr, F.~Vila, S.~D. Dalosto, E.~L. Shirley, and Z.~H.
  Levine.
\newblock Optical to uv spectra and birefringence of ${\text{sio}}_{2}$ and
  ${\text{tio}}_{2}$: First-principles calculations with excitonic effects.
\newblock {\em Phys. Rev. B}, 78:205108, Nov 2008.

\bibitem{LeNguyen2020}
Tue~Minh Le~Nguyen, Vo~Van~Hoang, and Hang T.~T. Nguyen.
\newblock Structural evolution of free-standing 2d silicon carbide upon
  heating.
\newblock {\em The European Physical Journal D}, 74(6):108, Jun 2020.

\bibitem{li2010architecture}
Guoxing Li, Yuliang Li, Huibiao Liu, Yanbing Guo, Yongjun Li, and Daoben Zhu.
\newblock Architecture of graphdiyne nanoscale films.
\newblock {\em Chemical Communications}, 46(19):3256--3258, 2010.

\bibitem{li2019new}
Linyang Li, Xiangru Kong, and Fran{\c{c}}ois~M Peeters.
\newblock New nanoporous graphyne monolayer as nodal line semimetal: Double
  dirac points with an ultrahigh fermi velocity.
\newblock {\em Carbon}, 141:712--718, 2019.

\bibitem{li2017psi}
Xiaoyin Li, Qian Wang, and Puru Jena.
\newblock $\psi$-graphene: a new metallic allotrope of planar carbon with
  potential applications as anode materials for lithium-ion batteries.
\newblock {\em The journal of physical chemistry letters}, 8(14):3234--3241,
  2017.

\bibitem{li2020architecture}
Yingjie Li, Yangyang Li, Peng Lin, Jing Gu, Xiaojun He, Moxin Yu, Xiaoting
  Wang, Chuan Liu, and Chunxi Li.
\newblock Architecture and electrochemical performance of alkynyl-linked
  naphthyl carbon skeleton: Naphyne.
\newblock {\em ACS applied materials \& interfaces}, 12(29):33076--33082, 2020.

\bibitem{Lin2012}
S.~S. Lin.
\newblock Light-emitting two-dimensional ultrathin silicon carbide.
\newblock {\em The Journal of Physical Chemistry C}, 116(6):3951--3955, Feb
  2012.

\bibitem{lin2024alkali}
Yung-Chang Lin, Rika Matsumoto, Qiunan Liu, Pablo Sol{\'\i}s-Fern{\'a}ndez,
  Ming-Deng Siao, Po-Wen Chiu, Hiroki Ago, and Kazu Suenaga.
\newblock Alkali metal bilayer intercalation in graphene.
\newblock {\em Nature communications}, 15(1):425, 2024.

\bibitem{PhysRevX.7.021019}
Qihang Liu and Alex Zunger.
\newblock Predicted realization of cubic dirac fermion in quasi-one-dimensional
  transition-metal monochalcogenides.
\newblock {\em Phys. Rev. X}, 7:021019, May 2017.

\bibitem{Liu2021}
Tianyang Liu, Yu~Jing, and Yafei Li.
\newblock Two-dimensional biphenylene: A graphene allotrope with superior
  activity toward electrochemical oxygen reduction reaction.
\newblock {\em The Journal of Physical Chemistry Letters}, 12(51):12230--12234,
  Dec 2021.

\bibitem{Long_2021}
Hui Long, Jianwei Hu, Xing Xie, Peiju Hu, Shaoxiong Wang, Minru Wen, Xin Zhang,
  Fugen Wu, and Huafeng Dong.
\newblock Sic siligraphene: a novel sic allotrope with wide tunable direct band
  gap and strong anisotropy.
\newblock {\em Journal of Physics D: Applied Physics}, 54(22):225102, mar 2021.

\bibitem{luder2015electronic}
Johann L{\'u}der, Monica de~Simone, Roberta Totani, Marcello Coreno, Cesare
  Grazioli, Biplab Sanyal, Olle Eriksson, Barbara Brena, and Carla Puglia.
\newblock The electronic characterization of biphenylene—experimental and
  theoretical insights from core and valence level spectroscopy.
\newblock {\em The Journal of chemical physics}, 142(7), 2015.

\bibitem{Luo2024}
Yi~Luo, Yiqiang He, Yunfei Ding, Lijie Zuo, Chengyong Zhong, Yinchang Ma, and
  Minglei Sun.
\newblock Defective biphenylene as high-efficiency hydrogen evolution
  catalysts.
\newblock {\em Inorganic Chemistry}, 63(2):1136--1141, Jan 2024.

\bibitem{luo2021first}
Yi~Luo, Chongdan Ren, Yujing Xu, Jin Yu, Sake Wang, and Minglei Sun.
\newblock A first principles investigation on the structural, mechanical,
  electronic, and catalytic properties of biphenylene.
\newblock {\em Scientific reports}, 11(1):19008, 2021.

\bibitem{RevModPhys.84.1419}
Nicola Marzari, Arash~A. Mostofi, Jonathan~R. Yates, Ivo Souza, and David
  Vanderbilt.
\newblock Maximally localized wannier functions: Theory and applications.
\newblock {\em Rev. Mod. Phys.}, 84:1419--1475, Oct 2012.

\bibitem{PhysRevMaterials.6.014012}
Mohammad~Ali Mohebpour, Shobair~Mohammadi Mozvashi, Sahar~Izadi Vishkayi, and
  Meysam~Bagheri Tagani.
\newblock Transition from metal to semiconductor by semi-hydrogenation of
  borophene.
\newblock {\em Phys. Rev. Mater.}, 6:014012, Jan 2022.

\bibitem{PhysRevB.13.5188}
Hendrik~J. Monkhorst and James~D. Pack.
\newblock Special points for brillouin-zone integrations.
\newblock {\em Phys. Rev. B}, 13:5188--5192, Jun 1976.

\bibitem{PhysRevLett.115.115501}
E.~Mostaani, N.~D. Drummond, and V.~I. Fal'ko.
\newblock Quantum monte carlo calculation of the binding energy of bilayer
  graphene.
\newblock {\em Phys. Rev. Lett.}, 115:115501, Sep 2015.

\bibitem{MOSTOFI20142309}
Arash~A. Mostofi, Jonathan~R. Yates, Giovanni Pizzi, Young-Su Lee, Ivo Souza,
  David Vanderbilt, and Nicola Marzari.
\newblock An updated version of wannier90: A tool for obtaining
  maximally-localised wannier functions.
\newblock {\em Computer Physics Communications}, 185(8):2309--2310, 2014.

\bibitem{PhysRevB.71.205214}
Nicolas Mounet and Nicola Marzari.
\newblock First-principles determination of the structural, vibrational and
  thermodynamic properties of diamond, graphite, and derivatives.
\newblock {\em Phys. Rev. B}, 71:205214, May 2005.

\bibitem{nose1984molecular}
Sh{\=u}ichi Nos{\'e}.
\newblock A molecular dynamics method for simulations in the canonical
  ensemble.
\newblock {\em Molecular physics}, 52(2):255--268, 1984.

\bibitem{10.1063/1.447334}
Shuichi Nosé.
\newblock {A unified formulation of the constant temperature molecular dynamics
  methods}.
\newblock {\em The Journal of Chemical Physics}, 81(1):511--519, 07 1984.

\bibitem{science.1102896}
K.~S. Novoselov, A.~K. Geim, S.~V. Morozov, D.~Jiang, Y.~Zhang, S.~V. Dubonos,
  I.~V. Grigorieva, and A.~A. Firsov.
\newblock Electric field effect in atomically thin carbon films.
\newblock {\em Science}, 306(5696):666--669, 2004.

\bibitem{RevModPhys.74.601}
Giovanni Onida, Lucia Reining, and Angel Rubio.
\newblock Electronic excitations: density-functional versus many-body
  green's-function approaches.
\newblock {\em Rev. Mod. Phys.}, 74:601--659, Jun 2002.

\bibitem{pan2021direct}
Qingyan Pan, Siqi Chen, Chenyu Wu, Feng Shao, Jing Sun, Lishui Sun, Zhaohui
  Zhang, Yixiao Man, Zhibo Li, Lixia He, et~al.
\newblock Direct synthesis of crystalline graphtetrayne—a new graphyne
  allotrope.
\newblock {\em CCS Chemistry}, 3(4):1368--1375, 2021.

\bibitem{PhysRevLett.77.3865}
John~P. Perdew, Kieron Burke, and Matthias Ernzerhof.
\newblock Generalized gradient approximation made simple.
\newblock {\em Phys. Rev. Lett.}, 77:3865--3868, Oct 1996.

\bibitem{D1NR07959J}
M.~L. Pereira, W.~F. da~Cunha, R.~T. de~Sousa, G.~D. Amvame~Nze, D.~S. Galvão,
  and L.~A. Ribeiro.
\newblock On the mechanical properties and fracture patterns of the
  nonbenzenoid carbon allotrope (biphenylene network): a reactive molecular
  dynamics study.
\newblock {\em Nanoscale}, 14:3200--3211, 2022.

\bibitem{PhysRevLett.130.076203}
C.~M. Polley, H.~Fedderwitz, T.~Balasubramanian, A.~A. Zakharov, R.~Yakimova,
  O.~B\"acke, J.~Ekman, S.~P. Dash, S.~Kubatkin, and S.~Lara-Avila.
\newblock Bottom-up growth of monolayer honeycomb sic.
\newblock {\em Phys. Rev. Lett.}, 130:076203, Feb 2023.

\bibitem{PhysRevB.62.4927}
Michael Rohlfing and Steven~G. Louie.
\newblock Electron-hole excitations and optical spectra from first principles.
\newblock {\em Phys. Rev. B}, 62:4927--4944, Aug 2000.

\bibitem{SHAHROKHI2016377}
Masoud Shahrokhi.
\newblock Quasi-particle energies and optical excitations of zns monolayer
  honeycomb structure.
\newblock {\em Applied Surface Science}, 390:377--384, 2016.

\bibitem{SHAHROKHI20171185}
Masoud Shahrokhi and Céline Leonard.
\newblock Tuning the band gap and optical spectra of silicon-doped graphene:
  Many-body effects and excitonic states.
\newblock {\em Journal of Alloys and Compounds}, 693:1185--1196, 2017.

\bibitem{sharma2014pentahexoctite}
Babu~Ram Sharma, Aaditya Manjanath, and Abhishek~K Singh.
\newblock pentahexoctite: A new two-dimensional allotrope of carbon.
\newblock {\em Scientific reports}, 4(1):7164, 2014.

\bibitem{PhysRevB.108.235311}
Arushi Singh, Vikram Mahamiya, and Alok Shukla.
\newblock Defect-driven tunable electronic and optical properties of
  two-dimensional silicon carbide.
\newblock {\em Phys. Rev. B}, 108:235311, Dec 2023.

\bibitem{song2013graphenylene}
Qi~Song, Bing Wang, Ke~Deng, Xinliang Feng, Manfred Wagner, Julian~D Gale,
  Klaus M{\'u}llen, and Linjie Zhi.
\newblock Graphenylene, a unique two-dimensional carbon network with
  nondelocalized cyclohexatriene units.
\newblock {\em Journal of Materials Chemistry C}, 1(1):38--41, 2013.

\bibitem{Strinati1988}
G.~Strinati.
\newblock Application of the green's functions method to the study of the
  optical properties of semiconductors.
\newblock {\em La Rivista del Nuovo Cimento (1978-1999)}, 11(12):1--86, Dec
  1988.

\bibitem{Susi2017}
Toma Susi, Viera Sk{\'a}kalov{\'a}, Andreas Mittelberger, Peter Kotrusz, Martin
  Hulman, Timothy~J. Pennycook, Clemens Mangler, Jani Kotakoski, and Jannik~C.
  Meyer.
\newblock Computational insights and the observation of sic nanograin assembly:
  towards 2d silicon carbide.
\newblock {\em Scientific Reports}, 7(1):4399, Jun 2017.

\bibitem{PhysRevB.107.085114}
Meysam~Bagheri Tagani.
\newblock ${\mathrm{si}}_{9}{\mathrm{c}}_{15}$ monolayer: A silicon carbide
  allotrope with remarkable physical properties.
\newblock {\em Phys. Rev. B}, 107:085114, Feb 2023.

\bibitem{PhysRevLett.84.1716}
H.~Terrones, M.~Terrones, E.~Hern\'andez, N.~Grobert, J-C. Charlier, and P.~M.
  Ajayan.
\newblock New metallic allotropes of planar and tubular carbon.
\newblock {\em Phys. Rev. Lett.}, 84:1716--1719, Feb 2000.

\bibitem{Togo_2023}
Atsushi Togo, Laurent Chaput, Terumasa Tadano, and Isao Tanaka.
\newblock Implementation strategies in phonopy and phono3py.
\newblock {\em Journal of Physics: Condensed Matter}, 35(35):353001, jun 2023.

\bibitem{Tromer2023}
Raphael~M. Tromer, Marcelo~L. Pereira~J{\'u}nior, Kleuton~A. L.~Lima,
  Alexandre~F. Fonseca, Luciano~R. da~Silva, Douglas~S. Galv{\~a}o, and
  Luiz.~A. Ribeiro~Junior.
\newblock Mechanical, electronic, and optical properties of 8-16-4 graphyne: A
  2d carbon allotrope with dirac cones.
\newblock {\em The Journal of Physical Chemistry C}, 127(25):12226--12234, Jun
  2023.

\bibitem{VANHOANG2019236}
Vo~{Van Hoang}, Nguyen {Hoang Giang}, To~{Quy Dong}, and Tran {Thi Thu Hanh}.
\newblock Tetra-sic – new allotrope of 2d silicon carbide.
\newblock {\em Computational Materials Science}, 162:236--244, 2019.

\bibitem{VEERAVENKATA2021893}
Harish~P. Veeravenkata and Ankit Jain.
\newblock Density functional theory driven phononic thermal conductivity
  prediction of biphenylene: A comparison with graphene.
\newblock {\em Carbon}, 183:893--898, 2021.

\bibitem{wang2015phagraphene}
Zhenhai Wang, Xiang-Feng Zhou, Xiaoming Zhang, Qiang Zhu, Huafeng Dong, Mingwen
  Zhao, and Artem~R Oganov.
\newblock Phagraphene: a low-energy graphene allotrope composed of 5--6--7
  carbon rings with distorted dirac cones.
\newblock {\em Nano letters}, 15(9):6182--6186, 2015.

\bibitem{WEI2021159201}
Qun Wei, Ying Yang, Guang Yang, and Xihong Peng.
\newblock New stable two dimensional silicon carbide nanosheets.
\newblock {\em Journal of Alloys and Compounds}, 868:159201, 2021.

\bibitem{PhysRevLett.103.186802}
Li~Yang, Jack Deslippe, Cheol-Hwan Park, Marvin~L. Cohen, and Steven~G. Louie.
\newblock Excitonic effects on the optical response of graphene and bilayer
  graphene.
\newblock {\em Phys. Rev. Lett.}, 103:186802, Oct 2009.

\bibitem{D1TC04154A}
Pei Zhang, Tao Ouyang, Chao Tang, Chaoyu He, Jin Li, Chunxiao Zhang, Ming Hu,
  and Jianxin Zhong.
\newblock The intrinsic thermal transport properties of the biphenylene network
  and the influence of hydrogenation: a first-principles study.
\newblock {\em J. Mater. Chem. C}, 9:16945--16951, 2021.

\bibitem{zhang2021semimetallic}
Wei Zhang, Changchun Chai, Qingyang Fan, Yanxing Song, Yuqian Liu, Yintang
  Yang, Minglei Sun, and Udo Schwingenschlogl.
\newblock Semimetallic 2d alkynyl carbon materials with distorted type i dirac
  cones.
\newblock {\em The Journal of Physical Chemistry C}, 125(32):18022--18030,
  2021.

\bibitem{C9NR08755A}
Liujiang Zhou, Huilong Dong, and Sergei Tretiak.
\newblock Recent advances of novel ultrathin two-dimensional silicon carbides
  from a theoretical perspective.
\newblock {\em Nanoscale}, 12:4269--4282, 2020.

\end{thebibliography}

\end{document}